\documentclass[prc, showpacs, preprintnumbers,amsmath,amssymb]{revtex4}
\usepackage{amsmath}
\usepackage{amssymb}
\usepackage{color}
\usepackage{mathrsfs}
\usepackage{graphics}
\usepackage{graphicx}
\usepackage{dcolumn}
\usepackage{bm}

\begin{document}
\title{Differential Neutrino Rates and Emissivities from the Plasma Process in 
Astrophysical Systems} 

\author{Sa{\v s}a Ratkovi\'c}
\email{ratkovic@tonic.physics.sunysb.edu}

\author{Sharada Iyer Dutta}
\email{iyers@neutrino.ess.sunysb.edu}

\author{Madappa Prakash}
\email{prakash@snare.physics.sunysb.edu}

\affiliation{Department of Physics \& Astronomy, \\
        State University of New York at Stony Brook, \\
        Stony Brook, NY 11794-3800, USA}


\date{\today}

\begin{abstract}

The differential rates and emissivities of neutrino pairs from an
equilibrium plasma are calculated for the wide range of density and
temperature encountered in astrophysical systems. New analytical
expressions are derived for the differential emissivities which yield
total emissivities in full agreement with those previously calculated.
The photon and plasmon pair production and absorption kernels in the
source term of the Boltzmann equation for neutrino transport are
provided.  The appropriate Legendre coefficients of these kernels, in
forms suitable for multi-group flux-limited diffusion schemes are also
computed.

\end{abstract}
\smallskip
\maketitle


\section{Introduction}
\label{sec:Introduction}

A wide range of astrophysical phenomena are significantly influenced
by weak interaction processes that involve the emission or absorption
of neutrinos in matter at high density and/or temperature.  Examples
include neutrino
energy loss in degenerate helium cores of red giant stars
\cite{Gross,Raffelt(2000)}, cooling in pre-white dwarf interiors
\cite{O'BrienKawaler(2000)}, the short- and long-term cooling of neutron
stars \cite{Prak02,Yak02}, the deflagration stages of white dwarfs
which may lead to type Ia supernovae \cite{Wolfgang,Iwamoto},
explosive stages of type II (core-collapse) supernovae \cite{Bur00},
and thermal emission in accretion disks of gamma-ray bursters
\cite{Matteo,Kohri}. (The selected references contain more complete
references to prior and ongoing work.)  Depending on the density and
temperature of ambient matter, the emission of neutrinos via both
neutral and charged current interactions is an important energy-loss
mechanism, while scattering and absorption of neutrinos serve to
deposit energy into matter.  In recent years, it has been realized
that neutrino cross-sections in matter and the implementation of
accurate neutrino transport are critical to understand the precise
mechanisms which trigger explosive events.  A sterling example is
provided by gravitational-core-collapse supernova simulations in which
a strong coupling between neutrino transport and hydrodynamics (in
many cases supplemented by convection, rotation, and magnetic fields)
has been realized \cite{Trans}.

Neutrino transport in the supernova environment is described by a
Boltzmann transport equation, which is a nonlinear integro-partial
differential equation that describes the time rate of change of the
neutrino distribution function $f$.  Advances made to date in the
numerical solution of this equation in the supernova context may be
found in Refs.~\cite{Trans}.  Historically, multigroup methods (in
which the equation is discretized in energy groups) have involved the
use of moment equations.  When the temporal derivative of the first
order deviation in $f$ is set to zero, a diffusion equation is
obtained, but this cannot adequately handle the free-streaming regime
at low densities.  Flux limiting schemes have been used to bridge the
diffusive and free-streaming regimes, but these are somewhat arbitrary
and calibrations vary depending upon neutrino opacities and dynamics.
In addition, there is a coupling of different neutrino-energy groups,
especially because of neutrino-electron scatterings which involve
large energy transfers.  An additional complication in supernovae is
that the approach to thermal and chemical equilibrium, and the passage
from diffusive flow to free streaming, require careful treatment.
Even with modern parallel supercomputers, it is necessary to integrate
the Boltzmann equation over solid angles to reduce the dimensionality
of the problem, with a corresponding loss of information about the
neutrino angular distribution function.  This could be important in
regimes in which neutrino-driven convection, a 3-D phenomenon, is
occuring.

The basic microphysical inputs of accurate neutrino transport coupled
in hydrodynamical situations are the differential production and
absorption rates and their associated emissivities.  The precise forms
in which such inputs are required for multienergy treatment of
neutrinos is detailed in Ref. \cite{BRUENN1} (see also
Ref. \cite{BT02} in which other current developments are summarized).
The objective of this work is to make available differential rates and
emissivities for the thermal production of neutrino pairs from a
plasma ($\gamma^* \rightarrow \nu + \bar\nu $). Similar quantities for
the photoneutrino process, $e + \gamma \rightarrow e + \nu + \bar\nu$,
will be presented separately.

We wish to note that the total neutrino pair emissivity from the
plasma process has been investigated in several prior
works \cite{BRAATEN1,BEAUDET1,BEAUDET2,DICUS1,BOND1,SCHINDER1}, and a
comprehensive treatment of this process valid over a wide range of
density and temperature can be found in Ref. \cite{BRAATEN1}.  (The
results of this latter reference are used extensively in this work.)
However, in prior works in which total emissivities were computed 
the energy and angular dependences of the
emitted neutrinos were eliminated with the help of Lenard's identity:
\begin{equation}
\int \frac {d^3q_1}{2E_1}   \frac {d^3q_2}{2E_2} 
\delta^4(K-Q_1-Q_2 ) Q_1^\mu Q_2^\nu = \frac {\pi}{24} 
\Theta(K^2) (2 K^\mu K^\nu + K^2 g^{\mu\nu})  \,,  
\label{Lenard}
\end{equation}  
where $Q_1^{\mu}=(E_1,{\bf q_1})$ and $Q_2^{\mu}=(E_2,{\bf q_2})$ are
the 4-momenta of the outgoing neutrinos, and $K^\mu=(\omega,{\bf k})$
is the 4-momentum of the massive photon in the medium.  While the use
of Eq. (\ref{Lenard}) simplifies considerably the calculation of the
total emissivity, differential information about the neutrinos is
entirely lost.  In addition, calculations of differential rates and
emissivities entail the calculation of the relevant squared matrix
elements, which was hitherto bypassed in obtaining the total rates and
emissivities.

Section \ref {sec:Plasmon} is devoted to obtaining the
squared matrix elements which are employed in the calculation of the
differential rates and emissivities discussed in detail in Sec. \ref
{sec:DIFFEMI}.  A check of these results is provided by comparing the
total emissivities obtained by both analytical and numerical
integrations of the differential emissivities.  The results for the
total emissivities reported in Ref. \cite{BRAATEN1} are used as an
additional check of our results. Qualitative and quantitative
discussions of the numerical results obtained here are presented in
Sec. \ref{sec:SNR}. Sec. \ref{sec:PKER} contains
production and absorption kernels in forms suitable for detailed
calculations of neutrino transport. In order to assess the relative
importance of neutrino pair production from the plasma process, a
comparison with competing processes is made in
Sec. \ref{sec:Comparison} over a wide range of density and
temperature.  A summary of this work is provided in
Sec. \ref{sec:summary}. For completeness, a brief description of 
the standard model effective coupling is included in Appendix
\ref{sec:EFC}.  Except when presenting numerical results, we use units
in which $\hbar = c = k_B =1$.
\begin{figure}[h!]
\begin{picture}(220, 150)(0, 0)
\put(000,000){\includegraphics[width=0.25\textwidth]{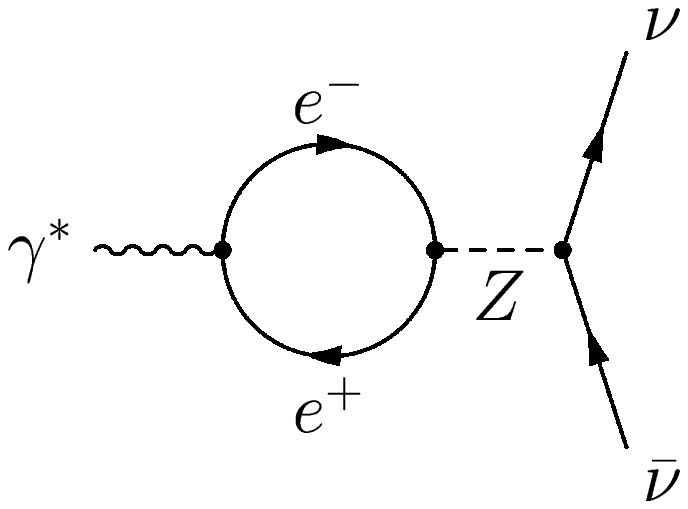}}
\put(110,000){\includegraphics[width=0.25\textwidth]{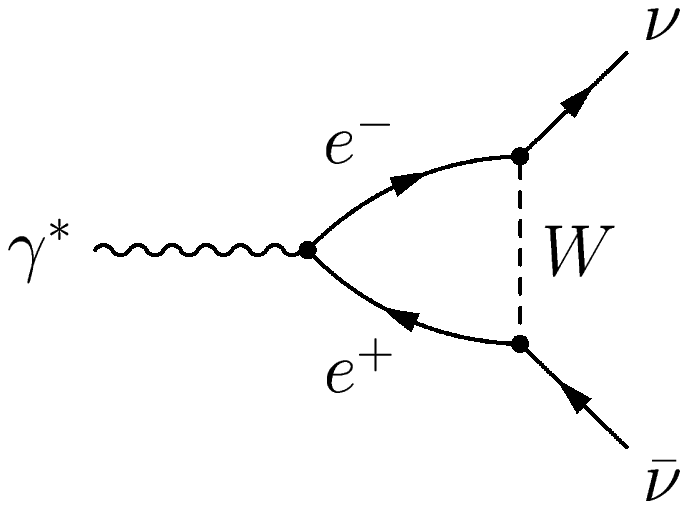}}
\end{picture}
\caption{Leading order Feynman diagrams describing the emission of a
neutrino pair from the plasma process. The charged current process
in which the $W$-boson is exchanged produces only $\nu_e \bar{\nu_e}$,
while that in which the neutral $Z$-boson is exchanged results in pairs of all
three neutrino ($e,~\mu,~{\rm and}~\tau$) flavors.}
\label{FEYNMAN}
\label{pair}
\label{photo}
\label{plasmon}
\end{figure}


\section{Photon and Plasmon Decay}
\label{sec:Plasmon}

Our discussion will be restricted to the case of an equilibrium plasma
in which the net negative electric charge of electrons and positrons
is cancelled by a uniform positively charged background of protons,
alpha particles, and heavier ions.  The equation of state and the
phase structure of matter, and the abundances of the various
constituents including those of dripped neutrons at sub-nuclear
densities are determined by the minimization of free energy.

As is well known, $e^+e^-$ pairs in a plasma cause the photon to
acquire an effective mass $K^2=\Pi\ne 0$, which arises from
electromagnetic interactions (cf. Ref. \cite{BRAATEN1} and references
therein).  Therefore, we can consider the photon to be a massive
spin--1 particle that couples to the $\nu\bar{\nu}$ pair through the
two one-loop diagrams shown in Fig.  \ref{plasmon}. The channel in
which the exchange of the $Z$-boson occurs can produce any of the
three species ($\nu_{e,\mu,\tau}$) of neutrinos and their
anti-particles, whereas the channel in which the $W$-boson is
exchanged, only the $\nu_e \bar \nu_e$ pair is produced.

\subsection{Effective Coupling}
\label{subsec:EPC}

At the low energies of interest here, the matrix element for neutrino
pair production from the plasma is given by
\begin{equation}
  \mathcal{M}=\epsilon^\mu(K) \Gamma_{\mu\nu}(K) \biggl[
  \bar{u}_1(Q_1) \gamma^\nu (1-\gamma_5) v_2(Q_2) \biggr] \,,
\end{equation}
where $\epsilon^\mu$ is the photon's polarization,
$\Gamma_{\mu\nu}(K)$ is the effective vertex tensor, $u_1$ and $v_2$
are the wave functions of the outgoing neutrinos, and the Dirac
$\gamma$-matrices have their usual meaning.  The 4-momenta $K,~Q_1$
and $Q_2$ are defined earlier in connection with Eq. (\ref{Lenard}).
The effective vertex can be expressed as
\begin{eqnarray}
\label{Gamma}
  \Gamma_{\mu\nu}(K) &\equiv& \frac{eG_FT}{\sqrt{2}}
  \mathbf{Tr}\bigg[\sum_{l=-\infty}^{+\infty}\int
  \frac{d^3p}{(2\pi)^3} \,\, \gamma_\mu\frac{1}{\not\! P+\not \! K - m_e} 
  \gamma_\nu \left(C_V^f-C_A^f\gamma_5\right)\frac{1}{\not\! P-m_e}\bigg] \,,
\end{eqnarray}
where $e$ and $m_e$ are the charge and mass of the electron, $G_F$ is
Fermi's weak interaction coupling constant, and $T$ is the temperature.
$C_V^f$ and $C_A^f$ are the vector and axial couplings for a
neutrino of flavor $f$. The trace is taken over the spin states and
$\sum_l$ denotes the sum over the Matsubara frequencies
\begin{eqnarray}
P_0=(2l+1)\pi T i +\mu_e
\end{eqnarray}
of the $e^+e^-$ loop in a heat bath with chemical potential $\mu_e$.
The $e^-$ and $e^+$ loop momenta are denoted by $P_1^\mu=(E_{e1}, {\bf
p_1})\equiv P^\mu=(E_{e}, {\bf p}) $ and $P_2^\mu=(E_{e2}, {\bf
p_2})=P^\mu - K^\mu$, respectively.  The $\Gamma_{\mu\nu}(K)$ in
Eq.~(\ref{Gamma}) satisfies current conservation, {\it i.e.,}
$K^\mu\Gamma_{\mu\nu}(K)=0$.
It is advantageous to decompose the vertex tensor $\Gamma_{\mu\nu}$
into its vector and axial parts:
\begin{eqnarray}\label{GAMMA1}
\Gamma_{\mu\nu}(K)&=&V_{\mu\nu}(K)+iA_{\mu\nu}(K)\,,\nonumber\\
V_{\mu\nu}(K)&\equiv&4\frac{eG_FTC_V^f}{\sqrt{2}} 
\sum_l\int\frac{d^3p}{(2\pi)^3} \,\,\bigg [
        \frac{2P_\mu P_\nu + g_{\mu\nu} (m_e^2-P^2-K\cdot P) + 
K_\mu P_\nu + K_\nu P_\mu} {(P^2-m_e^2)((P+K)^2-m_e^2)} \bigg]\,, \nonumber \\
A_{\mu\nu}(K) 
&\equiv&4\frac{eG_FTC_A^f}{\sqrt{2}}\sum_l\int\frac{d^3p}{(2\pi)^3}\,\,
        \frac{\varepsilon_{\mu\nu\alpha\beta}P^\alpha K^\beta}
        {(P^2-m_e^2)((P+K)^2-m_e^2)} \,. 
\end{eqnarray}
Summing over the Matsubara frequencies
and using contour integration, $V_{\mu\nu}(K)$ and $A_{\mu\nu}(K)$ can
be expressed as
 \begin{widetext}
\begin{eqnarray}\label{VA}
V_{\mu\nu}(K)&=&4\frac{eG_FC_V^f}{\sqrt{2}(2\pi)^3}\int\frac{d^3p}{2E_e} \,\,
\frac{(K_\mu P_\nu + K_\nu P_\mu)(K\cdot P) - K^2 P_\mu P_\nu 
- g_{\mu\nu} (K\cdot P)^2}
        {(K\cdot P)^2-(K^2)^2/4} 
        \Big(n_F(E_e, \mu_e, T)+\bar{n}_F(E_e, \mu_e, T)\Big) \,,\nonumber
\\ \nonumber \\
A_{\mu\nu}(K)&=&2\frac{eG_FC_A^f}{\sqrt{2}(2\pi)^3}\int\frac{d^3p}{2E_e} \,\,
\varepsilon_{\mu\nu\alpha\beta}
        \frac{K^2 (P^\alpha K^\beta)}
        {(K\cdot P)^2-(K^2)^2/4}
        \Big(n_F(E_e, \mu_e, T)-\bar{n}_F(E_e, \mu_e, T)\Big) \,.  
\end{eqnarray}
\end{widetext} 
Above, $n_F$ and $\bar{n}_F$ are the Fermi-Dirac distribution functions for
the electrons and positrons:
\begin{eqnarray}
n_F(E_e, \mu_e, T)=\frac{1}{e^{\frac{E_e-\mu_e}{T}}+1} \qquad {\rm{and}} \qquad
\bar{n}_F(E_e, \mu_e, T)=\frac{1}{e^{\frac{E_e+\mu_e}{T}}+1} \, 
\label{FERMI}
\end{eqnarray}
and $E_e = {\sqrt {p^2+m_e^2} }$ is the electron energy.  Note that
the denominators in Eq.~(\ref{VA}) give rise to higher than
first corrections in the fine structure constant
$\alpha = e^2/4\pi$, as can be seen by the expansion
\begin{eqnarray} 
\label{expansion}
        \frac{1}{(K\cdot P)^2-(K^2)^2/4}&=&\frac{1}{(K\cdot P)^2}+\mathcal{O}
        \left(\frac{(K^2)^2}{(K\cdot P)^2}\right)+\cdots \nonumber \\
        &=&\frac{1}{(K\cdot P)^2}+\mathcal{O}(\alpha^2)+\cdots
\end{eqnarray} 
The $\mathcal{O}(\alpha^2)$ term would allow the unphysical decay of
the plasmon into $e^+e^-$ pair through the imaginary part of the pole.
The prevention of undesired $\gamma^*\rightarrow e^+e^-$ decays
entails a calculation that includes higher order diagrams which serve
to increase the $e^+e^-$ pair mass and block such unphysical decays as
explained in Ref. \cite{BRAATEN1}.  Thus, to $\mathcal{O}(\alpha)$, it
is necessary to drop all but the first term on the right hand side of
Eq.~(\ref{expansion}).

\subsection{Polarization Functions}
\label{subsec:Psf}
In order to study the individual contributions of the different 
polarizations, it is convenient to express the vector part of
$\Gamma_{\mu\nu}$ in terms of its longitudinal and transverse
components. (Following Ref. \cite{BRAATEN1}, we will utilize the term
``photon'' for the transverse component and ``plasmon'' for the
longitudinal component.) The convention for these operators is adopted
from Refs.~ \cite{KAPUSTA,LEBELLAC}. The axial part must be treated
separately as it does not lend itself to such a decomposition.  Since
the polarization tensor is proportional to $V_{\mu\nu}$, {\it i.e.,}
\begin{eqnarray}
V_{\mu\nu}(K)= \frac{G_F C_V^f}{\sqrt{2} e} \Pi_{\mu\nu}(K) \,,
\end{eqnarray}
the scalar polarization functions $G(\omega_T,k)$ and $F(\omega_L,k)$ can be
defined through the relation
\begin{eqnarray}
        \Pi^{\mu\nu}(K)&=&G(\omega_T,k)P_T^{\mu\nu}+F(\omega_L,k)P_L^{\mu\nu} \,,
\end{eqnarray}
where the tensor structure of $ \Pi^{\mu\nu}$ is captured in the
polarization tensors $P_L^{\mu\nu}$ and $P_T^{\mu\nu}$. 
The components of 
the transverse and longitudinal polarization tensors are 
\begin{eqnarray}
P_T^{\mu\nu}&=&\left\{ \begin{array}{ll}
      0& \textrm{for } \mu \textrm{ or }\nu=0\\
      \delta^{ij}-\frac{k^ik^j}{k^2}&i,j=1, 2, 3
\end{array}
\right. 
\end{eqnarray}
\begin{eqnarray}
P_L^{\mu\nu}&=&\frac{K^\mu K^\nu}{K^2}-g^{\mu\nu}-P_T^{\mu\nu} \,.
\end{eqnarray}
The polarization tensors satisfy the properties
\begin{eqnarray}
P_T^{\mu\rho}P_{L\rho\nu}&=&0 \,, \quad
P_T^{\mu\rho}P_{T\rho\nu} = -P_{T\nu}^\mu \nonumber\\
P_L^{\mu\rho}P_{L\rho\nu}&=&-P_{L\nu}^\mu\,, \quad 
P_{L\mu}^\mu = -1 \,, \quad 
P_{T\mu}^\mu = -2 \,.
\end{eqnarray}
For completeness, we note that $P_L^{\mu\nu}$ can also be written
in the form of a tensor product \cite{BRAATEN1}:
\begin{eqnarray}
P_L^{\mu\nu}=\frac{k^2}{K^2}\,\left(1\quad\frac{k_0{\bf k}}{{k}^2}\right)\otimes\left( \begin{array}{c}
1\\ \\
\frac{k_0{\bf k}}{{k}^2}\\
\end{array}
\right)\,.
\end{eqnarray}
Utilizing these properties, we have the relations
\begin{eqnarray}\label{GF}
        G(\omega_T,k)&=&\frac{1}{2}P_T^{\mu\nu}\Pi_{\mu\nu}(K) \,,\nonumber \\
        F(\omega_L,k)&=&P_L^{\mu\nu}\Pi_{\mu\nu}(K) =
        \frac{K^2}{k^2}\Pi_{00}(K) \,.  
\end{eqnarray}
Performing the angular integrations in Eqs.~(\ref{GAMMA1}) and
(\ref{GF}), the scalar functions $G$ and $F$ take the form 
\begin{eqnarray}
\nonumber
        G(\omega_T,k)&=&\frac{e^2}{\pi^2}\int_0^\infty\frac{p^2dp}{E_e}\,\,
        \bigg[
        \frac{\omega_T^2}{k^2}
         - \frac{\omega_T^2-k^2}{2k^2}\frac{\omega_T}{vk^2}
        \ln{\left(\frac{\omega_T+vk}{\omega_T-vk}\right)}
        \bigg] (n_F+\bar{n}_F) \label{G}\,, \nonumber\\
        F(\omega_L,k)&=&\frac{\omega_L^2-k^2}{k^2}\frac{e^2}{\pi^2}\int_0^\infty
        \frac{p^2dp}{E_e} \,\, 
\bigg[\frac{\omega_L}{vk}\ln{\left(\frac{\omega_L+vk}{\omega_L-vk}\right)}
              -1 -\frac{\omega_L^2-k^2}{\omega_L^2-v^2k^2}
        \bigg](n_F+\bar{n}_F) \label{F} \,, 
\end{eqnarray}
where $v=p/E_e$ is the velocity of the electron.  These polarization
functions are compactly defined as
\begin{eqnarray}
\Pi_T(K)&\equiv& G(\omega_T,k) \,, \nonumber\\
\Pi_L(K)&\equiv&\Pi^{00}(K)\equiv ({k^2}/{K^2}) F(\omega_L,k) 
\label{POLS}
\end{eqnarray}
in Refs.~\cite{KAPUSTA,LEBELLAC}.  

In order to calculate the axial part, it is convenient to write
$A_{\mu\nu}$ in terms of a simple tensor and a scalar function, as was
done for $V_{\mu\nu}$ previously. We would like to keep manifest
Lorentz covariance in our expression, since frame fixing would require
special care later when the emissivities are calculated in the frame
of the heat bath.  We perform calculations in the Lorentz gauge
($\partial \cdot A=0$) which is suggested as a convenient choice by
the equations of motion in an effective theory of a massive spin--1
particle (other choices \cite{BRAATEN2} lead to the same
result). From the tensor structure of $A_{\mu\nu}$ in Eq. (\ref{VA}),
it is clear that only the transverse and longitudinal components of
the integral
\begin{eqnarray}
\int \frac {d^3p}{2E_e(2\pi)^3} 
\frac{P_\mu}{(K\cdot P)^2} \left(n_F-\bar{n}_F\right) 
\end{eqnarray}
contribute. The term ``transverse'' here refers to a 3--vector
transverse vector, {\it i.e.}, all three polarizations are 4--vector
transverse to $K^\mu$.  As a result, only the longitudinal component
survives the $d^3p$ integration (this is easy to see if we temporarily
rotate the photon momentum along the z--axis). The integral above is
then $\propto\epsilon^{(3)\mu}$, which in a general frame for a
massive vector particle of mass $M$ has the form \cite{fnote}
\begin{eqnarray}
  \epsilon^{(3)\mu}&=&\Big(\frac{k}{M},\, \frac{\omega {\bf k}}{kM}\Big)\, .
  \label{epsilon}
\end{eqnarray}
Using this property, it is clear that only the transverse
polarizations contribute to the emissivity. Thus $A_{\mu\nu}$ can be
recast as
\begin{eqnarray}
  A^{\mu\nu} &=& 
\frac{G_FC_A^f}{\sqrt{2}e}\varepsilon^{\mu\nu\alpha\beta}\epsilon^{(3)}_\alpha
  \frac{K_\beta}{\sqrt{K^2}}\Pi_A(K) \,,
  \label{Afinal}
\end{eqnarray}
where $\sqrt{K^2}=\sqrt{\Pi_T}$ is the mass of the vector particle and
the scalar function $\Pi_A(K)$ was chosen to coincide with that given
in Ref \cite{BRAATEN2}:
\begin{eqnarray}
  \Pi_A(K) &=& \frac{2\alpha}{\pi}\frac{K^2}{k}\int dp\,\frac{p^2}{E_e^2}
  \Big[ \frac{\omega_T}{2vk}\ln\left(\frac{\omega_T+vk}{\omega_T-vk}\right)-
  \frac{\omega_T^2-k^2}{\omega_T^2-v^2k^2}
  \Big]\big(n_F-\bar{n}_F\big)\,. 
\label{PiA}
\end{eqnarray}

\subsection{Squared Matrix Elements} 
\label{subsec:SME}

The squared matrix element for neutrino pair production in a plasma
can be computed using
\begin{eqnarray}
  |\mathcal{M}|^2 
  &=&\sum_{\lambda=1}^3\epsilon^{*(\lambda)}_{\mu}\epsilon^{(\lambda)}_{\nu}
  {\Gamma^*}^{\mu\alpha}{\Gamma}^{\nu\beta}L_{\alpha\beta} \,,
\end{eqnarray}
where the summation is over photon polarizations $\lambda$  
and the outgoing lepton tensor for neutrinos is 
\begin{eqnarray}
        L_{\alpha\beta}&=&8\big[Q_{1\alpha}Q_{2\beta}+Q_{1\beta}Q_{2\alpha}
        -g_{\alpha\beta}(Q_1\cdot Q_2) 
        +i\epsilon_{\alpha\beta\rho\sigma}Q_1^\rho Q_2^\sigma \big] \,.
\end{eqnarray}
The sum over polarizations can be performed using 
\begin{eqnarray*}
        \sum_{\lambda=1}^3{\epsilon^{*(\lambda)}}^{\mu}{\epsilon^{(\lambda)}}^\nu 
= - g^{\mu\nu} + \frac {K^\mu K^\nu}{K^2} \,.
\end{eqnarray*}
From the symmetry properties of
$\Gamma^{\mu\nu}=V^{\mu\nu}+iA^{\mu\nu}$ and
$L^{\mu\nu}=v^{\mu\nu}+ia^{\mu\nu}$, it is easy to deduce the
combinations that contribute to $|\mathcal{M}|^2$.  Explicitly,
\begin{eqnarray}
  |\mathcal{M}|^2 &=& -V^{\mu\alpha}V_{\mu}\,^{\beta}v_{\alpha\beta}
  -A^{\mu\alpha}A_{\mu}\,^{\beta}v_{\alpha\beta}
  +(V^{\mu\beta}A_{\mu}\,^{\alpha}
  -V^{\mu\alpha}A_{\mu}\,^{\beta})a_{\alpha\beta} \, ,
\end{eqnarray}
where the transversality of $\Gamma^{\mu\nu}$, namely that
$K^\mu\Gamma_{\mu\nu}=0$, was used.  The first term can be further
decomposed into its transverse and longitudinal components to yield
$|\mathcal{M}|_T^2$ and $|\mathcal{M}|_L^2$. The second term is purely
axial and will be denoted by $|\mathcal{M}|_A^2$, while the last term
is the mixed vector--axial term $|\mathcal{M}|_M^2$. It will be shown
later that this last term is purely transverse--axial (i.e. without
longitudinal contributions) and does not contribute to the total
emissivity and rate, but only to the differential emissivity and rate.

The squared matrix elements from the
transverse, longitudinal, axial, and mixed vector-axial channels are:
\begin{widetext}
\begin{eqnarray}
\langle|\mathcal{M}|^2\rangle_T
        &=& \frac{G_F^2 (C_V^f)^2}{\pi \alpha} \Pi_T^2(\omega_T, k)
        \bigg[E_1E_2 - \frac{({\bf k}\cdot{\bf q_1})({\bf k}\cdot{\bf
        q_2})}{k^2}\bigg]\,,
\nonumber\\
        \langle|\mathcal{M}|^2\rangle_L
        &=& 2\frac{G_F^2 (C_V^f)^2}{\pi \alpha} \Big(\frac{\omega_L^2-k^2}{k^2}\Big)^2 \Pi_L^2(\omega_L, k) \nonumber \\ && \times \quad
\bigg[\frac{(E_1\omega_L-{\bf q_1}\cdot{\bf k})(E_2\omega_L-{\bf q_2}\cdot{\bf k})}{\omega_L^2-k^2}
        + \frac{({\bf k}\cdot{\bf q_1})({\bf k}\cdot{\bf q_2})}{k^2}-
        \frac{E_1E_2+{\bf q_1}\cdot{\bf q_2}}{2}
        \bigg] \,, \nonumber \\
  \langle|\mathcal{M}|^2\rangle_A &=& 
\frac{G_F^2{(C_A^f)}^2}{\pi\alpha}{\Pi_A^2(\omega_T, k)}
\left[ E_1E_2-\frac{({\bf k}\cdot{\bf q_1})({\bf k}\cdot{\bf
  q_2})}{k^2}\right]\,,  \nonumber \\ 
\langle|\mathcal{M}|^2\rangle_M &=& 
2\frac{G_F^2 C_A^f C_V^f}{\pi\alpha} 
  \frac{{\Pi_A(\omega_T, k)}\Pi_T(\omega_T, k)}{k} \Big[ E_1({\bf k}\cdot{\bf q_2})-E_2({\bf k}
\cdot{\bf q_1})\Big]\,. 
\label{MM}
\end{eqnarray}
\end{widetext}
where the symbol $\langle \cdots \rangle$ indicates that the
appropriate spin averages and summations have been performed.  These
expressions are central to the differential and total emissivities
presented in this paper.

The main contribution to the total emissivity comes from the
transverse mode, the longitudinal component becoming comparable to the
transverse component only in the extremely degenerate limit
\cite{BRAATEN1,BEAUDET2}.  Contributions from the axial channel are
orders of magnitude smaller than the transverse and longitudinal
parts.  The mixed vector-axial total emissivity vanishes completely.
For all practical purposes, therefore, it is sufficient to consider
only the transverse and longitudinal parts, while the axial and mixed
vector-axial parts can be safely neglected.


\section{Differential and Total Emissivities}
\label{sec:DIFFEMI}
The total emissivity or the total energy carried away by the neutrino
pair per unit volume per unit time can be computed from
\begin{eqnarray}\label{QT}
        Q&=&\sum_{\epsilon}
        \int\frac{d^3k}{2\omega(2\pi)^3}\,\,Z(k)\,\,
        \frac{d^3q_1}{2E_1(2\pi)^3} \frac{d^3q_2}{2E_2(2\pi)^3}\,\,
        (E_1+E_2)\,\langle{|\mathcal{M}|}^2\rangle\, 
        n_B(\omega, T)\,(2\pi)^4 \, \delta^4(K-Q_1-Q_2) \,,
\end{eqnarray}
where $n_B(\omega,T)$ is the Bose-Einstein distribution function for
photons or plasmons and the sum is over appropriate polarizations. The
factor $Z(k)$ arises from the residue of the pole in the propagator:
\begin{eqnarray}
        \int d^4K~\delta(K^2-f(K)) = \int \frac {d^3k}
        {\left| 
2k_0- \frac {\partial f(k_0, {\bf k})} {\partial k_0}
          \right|}\quad \textrm{at} \quad k_0=\omega(\bf k) \,. 
\end{eqnarray} 
In vacuum, the free field solution has $f(k)=m^2$ and therefore the
denominator in the above expression is $2k_0({\bf k})=2\omega({\bf
k})$. In medium, various conventions for this correction can be found
in the literature \cite{BRAATEN1,BRAATEN2}. Here, $Z(k)$ is chosen such that
\begin{eqnarray}
        \frac{d^3k}{2\omega({\bf k})}Z(k)&=&
        \frac{d^3k}{\left|2\omega({\bf k})-\frac{\partial 
        f(\omega, {\bf k})}{\partial \omega}\right|} \,. 
\end{eqnarray}
Setting $f(\omega, {\bf k})= \Pi_T(\omega, {\bf k})$ in the case of
transverse polarization results in
\begin{eqnarray}
         Z_{T}(k) &=&\left|1-\frac{\partial 
            \Pi_{T}(\omega_T, {\bf k})}{\partial \omega_T^2}\right|^{-1} \,.
\end{eqnarray} 
A straightforward calculation then gives
\begin{widetext}
\begin{eqnarray}
        Z_T(k)&=&{\left|
            1-\frac{e^2}{\pi^2}\int_0^{\infty}\frac{p^2dp}{E_e}\,
            \left[
              \frac{3\omega_T^2-k^2(1+2v^2)}{2k^2(\omega_T^2-v^2k^2)}-
              \frac{3\omega_T^2-k^2}{4\omega_T^2k^3}
            \ln{\left(\frac{\omega_T+vk}{\omega_T-vk}\right)}
            \right]\,(n_F+\bar{n}_F)
          \right|}^{-1} \label{ZT}\,,
\end{eqnarray}
\end{widetext}
where $v\equiv p/E_e$.  Our choice   
\begin{widetext}
\begin{eqnarray}
        Z_L(k)&=&\frac{k^2}{\omega_L^2-k^2}{\left|
            \frac{e^2}{\pi^2}\int_0^{\infty}\frac{p^2dp}{E_e}\,
            \left[
              \frac{\omega_L^2-2v^2k^2+k^2}{(\omega_L^2-v^2k^2)^2}-
              \frac{1}{2\omega_L v k}\ln{\left(\frac{\omega_L+vk}{\omega_L-vk}\right)}
            \right]            \,(n_F+\bar{n}_F)
          \right|}^{-1} \label{ZL} \,
\end{eqnarray}
\end{widetext}
is slightly different from that used by Braaten~\cite{BRAATEN1}.
Specifically,
\begin{eqnarray}
Z_L=\omega_L^2/(\omega_L^2-k^2)\times Z_L^{BS}.
\end{eqnarray}
This difference results in some minor differences in the expressions
associated with the longitudinal emissivity.

\subsection{Transverse Differential Emissivity}
\label{sec:TDE}

Integrating Eq.~(\ref{QT}) over the 3-momentum of the photon, the
transverse differential emissivity is given by
\begin{widetext}  
\begin{eqnarray}
        \frac{d^3Q_T}{dE_1dE_2d\cos \theta}&=&
        \frac{G_F^2{\sum_f (C_V^f)}^2}{16\pi^4\alpha}
        \, Z_T(k) \, n_B(\omega_T, T) \, \Pi_T^2(\omega_T,k) \, 
                \frac{(E_1-E_2)^2+E_1 E_2 (1+ \cos \theta)}
        {E_1^2 + E_2^2 + 2E_1 E_2 \cos \theta}\nonumber\\&& \times \quad E_1^2 E_2^2 
(1- \cos \theta)\; \delta(\omega_T - E_1 - E_2) \,, 
\label{dQEET}
\end{eqnarray}
\end{widetext}
where $\theta$ is the angle between the two outgoing neutrinos of energy
$E_1$ and $E_2$. The dispersion relations are
\begin{eqnarray}
     {\omega}_T^2(k) &=& {k}^2+\Pi_T(\omega_T(k), k) \nonumber\\ 
 k &=& \sqrt{E_1^2 + E_2^2 + 2E_1 E_2 \cos \theta} \,.
\label{DISP}
\end{eqnarray}

Equation (\ref{dQEET}) is the basis upon which suitable forms of the
differential transverse emissivities can be obtained, since the energy
$\delta$-function can be used to advantage. For
example, integrating over one of the outgoing neutrino energies (say,
$E_2$) yields the differential emissivity as a function of the other
neutrino energy ($E_1=E$) and the angle $\theta$ between the pair
neutrinos. Explicitly,
\begin{eqnarray}
        \frac{d^2Q_T}{dE d\cos \theta}&=&
        \frac{G_F^2{\sum_f (C_V^f)}^2}{16\pi^4\alpha}
        \, Z_T(k) \, n_B(\omega_T, T) \, \Pi_T^2(\omega_T,k) \, 
	E^2 (\omega_T -E)^2 
        \frac{(2E-\omega_T)^2+E (\omega_T -E) (1+ \cos \theta)}
        {E^2 + (\omega_T -E)^2 + 2E (\omega_T -E) \cos \theta} \nonumber \\
        && \times \quad (1- \cos \theta) J_T(E, \theta) \,, 
\label{dQET} \\
        \nonumber 
\end{eqnarray}
where $J_T(E, \theta)$ is the Jacobian of transformation from $dE_2$
to $d(E_2-\omega_T(E_1, E_2))$ originating from the $\delta$-function
integration: 
\begin{eqnarray}
J_T(E, \theta) = \left|\frac{\partial E_2}{\partial(E_2-\omega_T(E_2))}
\right| = {\left|1-\frac{\partial \omega_T}{\partial E_2}\right|}^{-1} \,.
\label{JT} 
\end{eqnarray}
For computational ease, it is advantageous to cast the dispersion
relation in the form 
\begin{eqnarray}
  {\omega_T}&=&E+\frac{\Pi_T(\omega_T, k)}{2E(1-\cos \theta)}\,,
\label{OMEGAEPI}
\end{eqnarray}
which can be used to find $\omega_T(E, \theta)$ iteratively with
$\Pi_T\approx\omega_p^2$ (see Eq.~(\ref{WP})) used as a first guess.
In addition, the transcendental dispersion relation in  Eq.~(\ref{DISP})
has to be solved iteratively for every choice of $E$ and $\cos\theta$.
Using Eqs.~(\ref{G}) and (\ref{OMEGAEPI}), an explicit expression for
$J_T(E, \theta)$ can be derived. The result is
\begin{widetext}
\begin{eqnarray}
        J_T(E, \theta)&=&\left|\frac{(1-{\beta_T^{(E, \theta)}})\omega_T k^2}
                {k^2E(1-\cos\theta)-
                \Big[\omega_T(\omega_T^2-k^2)-E\omega_T^2(1-\cos\theta)
                \Big]{\beta_T^{(E, \theta)}}}\right| 
                \label{JET}\,,\nonumber \\ \nonumber \\
        {\beta_T^{(E, \theta)}}&=&\frac{e^2}{\pi^2}
         \int_0^\infty \frac{p^2dp}{E_e}\,\,
        \left[
        \frac{1}{k^2}+\frac{\omega_T^2-k^2}{2k^2(\omega_T^2-v^2 k^2)}
        +\frac{k^2-3\omega_T^2}{4v\omega_T k^3}
        \ln{\left(\frac{\omega_T+vk}{\omega_T-vk}\right)}
        \right](n_F+\bar{n}_F) \,.
\end{eqnarray}
\end{widetext}

Alternatively, the energy $\delta$-function in Eq.~(\ref{dQEET}) can also
be used to eliminate the angle between the two outgoing
neutrinos. This leads to
\begin{eqnarray}
        \frac{d^2Q_T}{dE_1 dE_2}&=&\frac{G_F^2{\sum_f(C_V^f)}^2}{16\pi^4\alpha}
        \, Z_T(k) \, n_B(\omega_T, T) \, \Pi_T^2(\omega_T, k) \, E_1 E_2 
        \bigg[E_1E_2 - \frac{(k^2+E_1^2-E_2^2)(k^2-E_1^2+E_2^2)}{4k^2} \bigg] 
        \nonumber \\
        && \times \quad  J_T(E_1, E_2) \,
        \Theta(4E_1E_2-\Pi_T) \,, 
 \label{dQEE}
\end{eqnarray}
where  $J_T(E_1, E_2)$ is the Jacobian resulting from the integration 
procedure. Its explicit form is given by  
\begin{eqnarray}
        J_T(E_1, E_2)&=&
        {\left|\frac{\partial\omega_T}{\partial{\cos\theta}}\right|}^{-1}
        =\frac{E_1+E_2}{E_1 E_2}\left|\frac{1-{\beta_T^{(E_1,
        E_2)}}}{1-\frac{\omega_T^2}{k^2}{\beta_T^{(E_1, E_2)}}}\right| 
\nonumber\,,
\label{JEE}
\end{eqnarray}
\begin{eqnarray}
        {\beta_T^{(E_1, E_2)}}&=&\frac{e^2}{\pi^2}
        \int_0^\infty \frac{p^2dp}{E_e}\,\bigg[
        \frac{1}{k^2}+\left(\frac{\omega_T^2}{k^2}-1\right)
        \frac{1}{2(\omega_T^2-v^2 k^2)} + 
        \frac{k^2-3\omega_T^2}{4v\omega_T k^3}
\ln{\left(\frac{\omega_T+vk}{\omega_T-vk}\right)}
        \bigg]\,(n_F+\bar{n}_F) \,.
\end{eqnarray}

\subsection{Longitudinal Differential Emissivity}
\label{sec:LDE}

The procedure outlined for the calculation of the transverse
differential emissivity can also be used to compute the longitudinal
differential emissivity.  However, there exists a difference between
the two cases. For the longitudinal component,  $k\leq
k_{max}$, so that the condition 
$E_1+E_2\leq k_{max}$ has to be satisfied. This can be
ensured by using the multiplicative factor
$\Theta(\mathcal{K} = k_{max}-E_1-E_2)$.  Utilizing the energy
$\delta$-function to perform the integration over the Euler angles, we
obtain
\begin{eqnarray}
        \frac{d^3Q_L}{dE_1dE_2d\cos \theta}&=&
     \frac{G_F^2{\sum_f (C_V^f)}^2}{16\pi^4\alpha}
        \, Z_L(k) \, n_B(\omega_L, T) \,
        (\omega_L^2-{k}^2)^2 \, E_1^3 E_2^3 
        \frac{1-\cos^2\theta} {k^2}
        \delta(\omega_L- E_1 - E_2)
        \Theta(\mathcal{K}) \label{dQLEET} \,,
\end{eqnarray}
where 
\begin{equation}
\omega^2_L(k) = \frac{\omega^2_L(k)}{k^2}\Pi_L(\omega_L(k), k) \,.
\label{LONG}
\end{equation}  

We first integrate over one of the neutrino energies to obtain the
differential emissivity depending on $E$ and $\theta$:
\begin{eqnarray}
        \frac{d^2Q_L}{dE\,d\cos\theta}&=&
   \frac{G_F^2{\sum_f (C_V^f)}^2}{16\pi^4\alpha}
 \, Z_L(k) \, n_B(\omega_L, T) \, (\omega_L^2-{k}^2)^2 \, E^3(\omega_L-E)^3 
        \frac{1-\cos^2\theta}{k^2} \Theta(\mathcal{K}) J_L(E,\theta)\,, 
\label{dQLET}\nonumber\\
    \nonumber \\
k&=&\sqrt{E^2 + (\omega_L-E)^2 + 2E(\omega_L-E) \cos \theta}\,, \\\nonumber \\
J_L(E, \theta)&=&{\left|
            \frac{\omega_L(1-\beta_L^{(E, \theta)})}
            {\omega_L(1-\beta_L^{(E,
\theta)})-(\omega_L-E(1-\cos\theta))
(1-\frac{\omega_L^2}{k^2}\beta_L^{(E, \theta)})}
          \right|} \,, \label{JLET} \nonumber\\ 
\beta_L^{(E, \theta)}&=&1+\frac{\omega_L^2-k^2}{k^2}\frac{e^2}{\pi^2}
        \int_0^\infty \frac{p^2dp}{E_e}\,
        \left[ \frac{2v^2k^2-\omega_L^2-k^2}{(\omega_L^2-v^2 k^2)^2}
     +\frac{1}{2v\omega_L k}\ln{\left(\frac{\omega_L+vk}{\omega_L-vk}\right)}
        \right] \,(n_F+\bar{n}_F) \,. 
\end{eqnarray}
On the other hand, performing the angular integration analogously to the
transverse case gives
\begin{eqnarray}
\frac{d^2Q_L}{dE_1dE_2}&=&\frac{G_F^2{\sum_f (C_V^f)}^2}{16\pi^4\alpha}
\, Z_L(k) \, n_B(\omega_L, T) \,
\frac{(\omega_L^2-{k}^2)^2 E_1^3 E_2^3 
(1-\cos^2\theta)}{E_1^2+E_2^2+2E_1E_2 \cos\theta} \nonumber\\ 
&& \times \quad J_L(E_1, E_2)\, \Theta(\mathcal{K}) \,
\Theta\Big(4E_1E_2-(\omega_L^2-k^2)\Big) \,,
        \label{dQLEE} \\
        \nonumber \\
J_L(E_1, E_2)&=&\frac{E_1+E_2}{E_1 E_2}\left|\frac{1-{\beta_L^{(E_1,
        E_2)}}}{1-\frac{\omega_L^2}{k^2}{\beta_L^{(E_1, E_2)}}}\right|
        \label{JLEE}\,\nonumber, \\\nonumber\\ \nonumber \\
\beta_L^{(E_1, E_2)}&=&1+\frac{\omega_L^2-k^2}{k^2}
        \frac{e^2}{\pi^2}
        \int_0^\infty \frac{p^2dp}{E_e}\,
        \left[ \frac{2v^2k^2-\omega_L^2-k^2}{(\omega_L^2-v^2 k^2)^2}
          +\frac{1}{2v\omega_L k}\ln{\left(\frac{\omega_L+vk}{\omega_L-vk}\right)}
        \right]\,(n_F+\bar{n}_F) \,,
\end{eqnarray}
where $\cos \theta=(k^2-E_1^2-E_2^2)/(2E_1E_2)$. The factor
$\Theta\Big(4E_1E_2-(\omega_L^2-k^2)\Big)$ in Eq.~(\ref{dQLEE}) accounts for
the fact that $|\cos\theta|$ is bounded by unity.

\subsection{Axial Differential Emissivity}
\label{sec:ADE}

Inserting $\langle|\mathcal{M}_A|^2\rangle$ from Eq.~(\ref{MM}) into 
Eq.~(\ref{QT}) and integrating over the photon momentum, 
the axial contribution to the differential emissivity is given by
\begin{widetext}  
\begin{eqnarray}
        \frac{d^3Q_A}{dE_1dE_2d\cos \theta}&=&
        \frac{G_F^2{\sum_f (C_A^f)}^2}{16\pi^4\alpha}
        \, Z_T(k) \, n_B(\omega_T, T) \,  \Pi_A^2(\omega_T, k) \,
        \frac{(E_1-E_2)^2+E_1 E_2 (1+ \cos \theta)}
        {E_1^2 + E_2^2 + 2E_1 E_2 \cos \theta}\nonumber\\ 
 && \times \quad \,E_1^2 E_2^2 \,
(1- \cos \theta)\; \delta(\omega_T- E_1 - E_2) \,.
\label{dQAEET}
\end{eqnarray}
\end{widetext}
Integrating over one of the neutrino energies yields
\begin{widetext}
\begin{eqnarray}
  \frac{d^2Q_A}{dE d\cos\theta} 
&=&\frac{G_F^2\sum_f{(C_A^f)}^2}{16\pi^4\alpha}
\, Z_T(k) \, n_B(\omega_T, T) \, \Pi_A^2(\omega_T, k) \,   
E^2(\omega_T-E)^2 (1 - \cos \theta)\nonumber 
\\&& \times \quad\,
\frac{(2E-\omega_T)^2+E (\omega_T -E) (1+ \cos \theta)}
        {E^2 + (\omega_T -E)^2 + 2E (\omega_T -E) \cos \theta} J_T(E, \theta)\,,
\label{dQAET}
\end{eqnarray}
where the residue factor $Z_T$ and the Jacobian $J_T(E, \theta)$ are
given by Eqs. (\ref{ZT}) and (\ref{JET}).

The corresponding expression for the differential emissivity as a
function of the two outgoing neutrino energies is
\begin{eqnarray}\label{dQAEE}
  \frac{d^2Q_A}{dE_1 dE_2}&=&\frac{G_F^2\sum_f{(C_A^f)}^2}{16\pi^4\alpha} 
\, Z_T(k) \, n_B(\omega_T, T) \, \Pi_A^2(\omega_T, k) \,
  E_1 E_2  \bigg[E_1E_2 - \frac{(k^2+E_1^2-E_2^2)(k^2-E_1^2+E_2^2)}{4k^2}
  \bigg] 
\nonumber \\
  &&\times\quad J_T(E_1, E_2) \, \Theta(4E_1E_2-\Pi_T) \,,
\end{eqnarray}
\end{widetext}
where $J_T(E_1, E_2)$ and $\Theta(4E_1E_2-\Pi_T)$ originate from
integrating over the $\delta$--function, and $J_T(E_1, E_2)$ is given
in Eq. (\ref{JEE}).

\subsection{Differential Emissivity from the Mixed Vector-Axial Channel}

The differential emissivities in this case are
\begin{widetext}
\begin{eqnarray}
\frac{d^3Q_M}{dE_1dE_2d\cos \theta}&=&
\frac{G_F^2{\sum_f C_A^f C_V^f}}{8\pi^4\alpha}
\, Z_T(k) \, n_B(\omega_T, T) \, 
\frac{\Pi_A(\omega_T, k) \Pi_T(\omega_T, k)}{k} \, E_1^2 E_2^2 (E_2-E_1)
\nonumber\\&& \times \quad \,
(1- \cos \theta)\,\delta(\omega_T- E_1 - E_2)  \label{dQMEET}\,,\\
\frac{d^2Q_M}{dE d\cos\theta}&=&\frac{G_F^2 
\sum_f C_A^f C_V^f}{8\pi^4\alpha} \, Z_T(k) \, n_B(\omega_T, T) \, 
\frac{ \Pi_A(\omega_T, k) \Pi_T(\omega_T, k)}{k} \, E^2(\omega_T-E)^2 
  (\omega_T-2E) \nonumber\\&& \times \quad \,(1 - \cos \theta)\,  J_T(E, \theta) 
\label{dQMET}\,,\\
  \frac{d^2Q_M}{dE_1 dE_2}&=&\frac{G_F^2 
\sum_f C_A^f C_V^f}{16\pi^4\alpha} \, Z_T(k) \, n_B(\omega_T, T) \,  
\frac{ \Pi_A(\omega_T, k) \Pi_T^2(\omega_T, k)}{k} \, E_1 E_2 (E_2-E_1)
   \nonumber \\&& \times \quad J_T(E_1, E_2)\, \Theta(4E_1E_2-\Pi_T) \label{dQMEE}\,.
\end{eqnarray}
\end{widetext}
Integrating over $(E, \theta)$ or $(E_1, E_2)$ confirms that the total
emissivity $Q_M=0$, a result that was anticipated from the $\nu_1
\leftrightarrow \nu_2$ antisymmetry of the squared matrix element in
Eq. (\ref{MM}).

The differential emissivities for the transverse component in
Eqs.~(\ref{dQET}) and (\ref{dQEE}), the longitudinal component in
Eqs.~(\ref{dQLET}) and (\ref{dQLEE}), the axial part in
Eqs. (\ref{dQAET}) and (\ref{dQAEE}), and the mixed axial-vector part
in Eqs. (\ref{dQMET}) and (\ref{dQMEE}) are among the principal new
results of this work.

\subsection{ Total Emissivity}
\label{sec:Emissivity}

The total emissivity can be obtained from Eq.~(\ref{QT}) by
performing integration over the entire phase space and summing
over all three outgoing $\nu$ flavors.  
Accounting for a factor of $2$ for the two
transverse photon polarizations, the transverse emissivity takes the form
\begin{widetext}
\begin{eqnarray}\label{QTTOTAL}
        Q_T&=&2\frac{G_F^2{\sum_f (C_V^f)}^2}{\pi\alpha}
        \int\frac{d^3k}{2\omega_T(2\pi)^3}\,\,Z_T(k)\,\,
        \frac{d^3q_1}{2E_1(2\pi)^3}\,\,
        \frac{d^3q_2}{2E_2(2\pi)^3} \,\,
        (E_1+E_2)n_B(\omega_T, T) (2\pi)^4 \delta^4(K-Q_1-Q_2) \nonumber \\
        && \times\quad \Pi_T^2(\omega_T, k) \,
        \left[E_1E_2-
        \frac{({\bf k}\cdot{\bf q}_1)({\bf k}\cdot{\bf q}_2)}{k^2}
        \right] \,, \nonumber \\
        &=&\frac{G_F^2{\sum_f (C_V^f)}^2}{48\pi^4\alpha}
        \int_0^\infty dk \,\, k^2 \, Z_T(k) \, n_B(\omega_T, T) \,
         (\omega_T^2- k^2)^3 \,.
\end{eqnarray}
\end{widetext}
Similarly, we obtain the longitudinal emissivity for single 
plasmon polarization as 
\begin{eqnarray}\label{QLTOTAL}
        Q_L&=&
        \frac{G_F^2{\sum_f (C_V^f)}^2}{96\pi^4\alpha}
        \int_0^{k_{max}} dk\,\, k^2 \, Z_L(k) \, n_B(\omega_L, T) \,  
         (\omega_L^2- k^2)^3 \,, 
\end{eqnarray}
where $k_{max}$ is the light-cone limit ({\it i.e.}, the maximum $k$
for which $k\le \omega_L$) for the longitudinal plasmon.  The results
in Eqs.~(\ref{QTTOTAL}) and (\ref{QLTOTAL}) agree with those of
Ref.~\cite{BRAATEN1} modulo slight differences in the conventions used
for $Z_L$.

Integration over the two outgoing neutrino momenta and angles in
Eqs. (\ref{MM}) and (\ref{QT}) yields the total axial emissivity
\begin{eqnarray}
Q_A &=& \label{QATOTAL}
\frac{G_F^2\sum_f{C_A^f}^2}{48\pi^4\alpha} 
\int dk\, k^2 \, Z_T(k)\,  n_B(\omega_T,T) \, \Pi_A^2(\omega_T, k) \, (\omega_T^2-k^2) \,,
\end{eqnarray}
which agrees with the result given previously in Ref. \cite{BRAATEN1}.
As noted earlier, the total mixed vector-axial 
emissivity $Q_M=0$. 

A cross--check of $Q_T,~Q_L$ and $Q_A$ is afforded by integrations of
their respective differential emissivities and by the results reported
in Ref. \cite{BRAATEN1}.  We present such checks in the following
section.

\section{Results and Discussion}
\label{sec:SNR}

For the most part, we will present our results for the total and
differential emissivities as a function of the mass density of protons
in the plasma, $\rho_BY_e = m_pn_e$, where $m_p$ is the proton mass,
$Y_e = n_e/n_B$ is the net electron fraction ($n_B$ is the baryon
number density) , and $n_e$ is net electron number density
\begin{eqnarray}
n_e(T, \mu_e)&=&\frac{1}{\pi^2}\int_0^\infty dp\,\,  p^2 \left(n_F -
\bar{n}_F \right) \,,
\label{MU}
\end{eqnarray}
which is simply the difference between the $e^-$ and $e^+$ number
densities. (The quantity  $\rho_BY_e$ is the same as $\rho/\mu_e$ used in prior
works including Ref. \cite{BRAATEN1}.)  Given $n_e$, this expression can be inverted to determine
the chemical potential $\mu_e$ at a given temperature.  In
connection with supernova simulations, however, it will be more
natural to examine emissivities as a function of the baryon mass
density $\rho_B = m_pn_B$. 

The differential and total emissivities require as
inputs the transverse and longitudinal dispersion relations
$\omega_{T,L}(k)$ in Eqs.~(\ref{DISP}) and (\ref{LONG}) , which are transcendental
relations that involve the polarization functions in Eq.~(\ref{POLS})
which are in turn expressed as integrals that involve
$\omega_{T,L}(k)$.  In addition, the residue factors $Z_{T,L}$ in
Eqs.~(\ref{ZT}) and (\ref{ZL}) also require $\omega_{T,L}(k)$ as
inputs.

With the help of the approximate expressions developed earlier in the
literature (see, for example, \cite{BRAATEN1,BEAUDET2} and
Sec.~\ref{sec:APP} below), we have calculated the exact dispersion
relations numerically by using iterative techniques.  Energy
integrations were performed by using standard Newton-Cotes algorithms.
Here an appropriate multiple of the temperature was used as the high energy
cutoff. Angular integrations were performed by employing Gauss-Legendre
quadrature with $N=16, 32$ and 64 points, all of which converged to the
same result.

\begin{figure}[ht]
\includegraphics[width=0.45\textwidth]{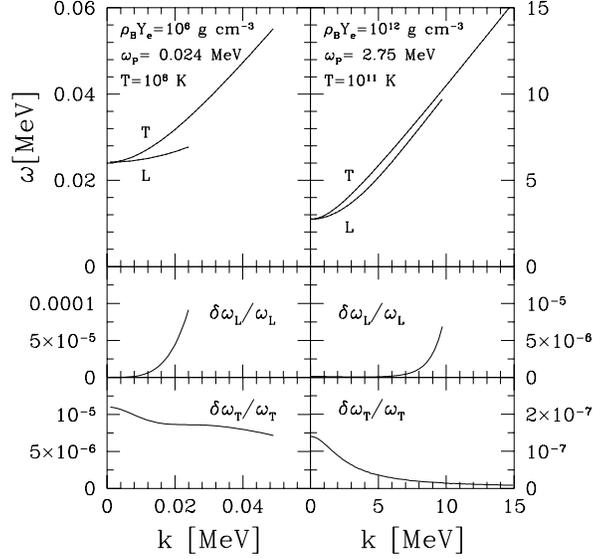}
\caption{Top panel: Dispersion relations in Eqs.~(\ref{DISP})
calculated numerically using the exact polarization functions in
Eqs.~(\ref{POLS}) and (\ref{LONG}) at the indicated temperatures and
densities.  The middle and bottom panels show the relative accuracies
of the longitudinal and transverse dispersion relations calculated
using the approximate polarization functions in Sec. \ref{sec:APP}. }
\label{approxexact} 
\end{figure}

Figure \ref{approxexact} shows the exact results of transverse and
longitudinal dispersion relations for temperatures $T=10^8$~K and
$T=10^{11}$~K, respectively. These calculations are numerically
cumbersome and time consuming, particularly in the case that
differential emissivities are needed. We therefore turn to assess the
extent to which the approximations developed in Ref. \cite{BRAATEN1}
reproduce the exact results over a wide range of temperature and
density.

\subsection{Approximations for $Z$,  $\Pi$, $J$ and $\beta$}
\label{sec:APP}

For completeness, we collect here the approximate relations developed
in Ref. \cite{BRAATEN1}.  These relations are particularly helpful in
developing new expressions that are needed in the calculation of
differential emissivities.  For the transverse mode, the residue
factor $Z_T$ and polarization function $\Pi_T$ are:
\begin{eqnarray} 
        Z_T(k)=\frac{2\omega_T^2(\omega_T^2-v_*^2k^2)}
                {3\omega_p^2\omega_T^2+(\omega_T^2+k^2)(\omega_T^2-v_*^2k^2)
                -2\omega_T^2(\omega_T^2-k^2)}\,,  \label{ZTA}
\end{eqnarray}
\begin{eqnarray} 
        \Pi_T(\omega_T,
                k)=\omega_p^2\frac{3}{2v_*^2}\left[\frac{\omega_T^2}{k^2} -
                  \frac{\omega_T^2-v_*^2k^2}{k^2}\frac{\omega_T}{2v_*k}
   \ln{\left(\frac{\omega_T+v_*k}{\omega_T-v_*k}\right)}\right] \,,
\label{PITA}
\end{eqnarray}
where
\begin{eqnarray}
        \omega_p^2&=&\frac{4\alpha}{\pi}\int_0^\infty dp\,\,\frac{p^2}{E_e}
                \left(1-\frac{v^2}{3}\right)(n_F+\bar{n}_F)\,, \label{WP} \\
        \omega_1^2&=&\frac{4\alpha}{\pi}\int_0^\infty dp\,\,\frac{p^2}{E_e}
                \left(\frac{5v^2}{3}-v^4\right)(n_F+\bar{n}_F)\,, \label{W1} \\
        v_*&=&\omega_1/\omega_p \,. \label{VS}
\end{eqnarray}
In Eq.~(\ref{WP}), $\omega_p$ is the plasma frequency -- the lower
limit for the value of plasmon mass (in this case \emph{transverse},
since $\Pi_T\ge\omega_p^2$). The quantities $\omega_1$ and $v_*$ are
auxiliary variables that render the expressions for $Z_Y$ and $\Pi_Y$
($Y=$ T or L) compact.

Equation (\ref{PITA}) can also be used to obtain approximate results for
Eqs.~(\ref{JET}) and (\ref{JEE}). For $J_T(E, \theta)$, we get
\begin{eqnarray} 
        J_T(E, \theta)= \left|
                \frac{\omega_T(1-{\tilde{\beta}_T^{(E, \theta)}})}
                {\omega_T(\frac{\omega_T^2}{k^2}-1)\tilde{\beta}_T^{(E,
                \theta)} + 
 E(1-\cos\theta)(1-\frac{\omega_T^2}{k^2}\tilde{\beta}_T^{(E, \theta)})}
                \right|  \label{JETA}
\end{eqnarray}
with
\begin{eqnarray}
\tilde{\beta}_T^{(E, \theta)}&\equiv& \frac{3}{4}\frac{\omega_p^2}{v_*^2k^2}
                \left[
3-\frac{3\omega_T^2-v_*^2k^2}{2v_*k\omega_T}\ln{\left(\frac{\omega_T+v_*k}{\omega_T-v_*k}\right)}
                \right]
\end{eqnarray}
and for $J_T(E_1, E_2)$, we obtain
\begin{eqnarray}
        J_T(E_1, E_2) = \frac{E_1+E_2}{E_1 E_2}
                \left|
\frac{1-\tilde{\beta}_T^{(E_1, E_2)}} 
{1-\frac{(E_1+E_2)^2}{k^2}\tilde{\beta}_T^{(E_1, E_2)}}
                \right|  \label{JEEA}
\end{eqnarray}
with 
\begin{eqnarray}
  \tilde{\beta}_T^{(E_1, E_2)} = \frac{9}{4}\frac{\omega_p^2}{v_*^2k^2}
                \left[
     1+\frac{1}{6}\left(\frac{v_*k}{\omega_T}-\frac{3\omega_T}{v_* k}\right)
                \ln{\left(\frac{\omega_T+v_*k}{\omega_T-v_*k}\right)}
                \right] \,, 
\end{eqnarray}
where $\omega_T=E_1+E_2$.

For the longitudinal mode, approximate expressions for $Z_L$ and
$\Pi_L$ are:
\begin{eqnarray} \label{ZLA}
        Z_L(k)&=&\frac{\omega_L^2}{\omega_L^2-k^2}\frac{2(\omega_L^2-v_*^2k^2)}
                {3\omega_p^2-(\omega_L^2-v_*^2k^2)}\,, \\ \nonumber \\
       \Pi_L(\omega_L, k)&=&\omega_p^2\,\,
              \frac{3}{v_*^2}\left[\frac{\omega_L}{2v_*k}
         \ln{\left(\frac{\omega_L+v_*k}{\omega_L-v_*k}\right)}-1\right] \,.
\label{PILA}
\end{eqnarray}
Compared to Ref. \cite{BRAATEN1}, the presence of the additional factor
$\omega_L^2/(\omega_L^2-k^2)$ in $Z_L$ stems from the differences in the
convention adopted for $Z_L$.  In this case, the Jacobians are:
\begin{eqnarray} 
        J_L(E, \theta)&=&{\left|{
        \frac{k^2(1-\tilde{\beta}_L^{(E,\theta)})}
        {k^2(1-\tilde{\beta}_L^{(E,\theta)})+\omega_L(\omega_L-E(1-\cos\theta))\tilde{\beta}_L^{(E,\theta)}}
        }\right|} \,, \\\label{JLETA}\nonumber\\
\tilde{\beta}_L^{(E, \theta)}&\equiv&\frac{3\omega_p^2}{2v_*^3}\,\,
        \bigg[
        \frac{3\omega_L}{2k^3} \ln{\left(\frac{\omega_L+v_*k}{\omega_L-v_*k}\right)}  
        -\frac{\omega_L^2}{k^2}\frac{v_*}{\omega_L^2-v_*^2k^2}-\frac{2v_*}{k^2}
        \bigg] \,, 
\end{eqnarray}
and
\begin{eqnarray} 
  J_L(E_1, E_2)&=&{\left|{
\frac{k^2}{E_1E_2\omega_L}
\frac{1-\tilde{\beta}_L^{(E_1, E_2)}}{\tilde{\beta}_L^{(E_1, E_2)}}
      }\right|} \,, 
\label{JLEEA}\\
        \tilde{\beta}_L^{(E_1, E_2)}&\equiv&\frac{3\omega_p^2}{2v_*^3}\,\,
        \bigg[ \frac{3\omega_L}{2k^3}
        \ln{\left(\frac{\omega_L+v_*k}{\omega_L-v_*k}\right)} 
        -\frac{\omega_L^2}{k^2}\frac{v_*}{\omega_L^2-v_*^2k^2}-\frac{2v_*}{k^2}
                \bigg] 
\end{eqnarray}
with $\omega_L=E_1+E_2$.
Ref. \cite{BRAATEN1} provides an approximate expression for $\Pi_A(\omega_T, k)$:
\begin{eqnarray}\label{PIAepp}
\Pi_A(\omega_T, k) &=& \omega_A k \frac{\omega_T^2-k^2}{\omega_T^2-v_*^2k^2}
        \frac{3\omega_p^2-2(\omega_T^2-k^2)}{\omega_p^2}
\end{eqnarray}
with
\begin{eqnarray}\label{Wa}
\omega_A &=& \frac{2\alpha}{\pi}\int_0^\infty\,dp\,\frac{p^2}{E^2}
        \big( 1 - \frac{2}{3} v^2 \big) (n_F-\bar{n}_F)\,.
\end{eqnarray}
We turn now to numerical results employing these approximations.  The
lower panels in Fig. \ref{approxexact} show the relative difference
between results calculated using these approximations and the exact,
but numerical, calculations of the dispersion relations.  At the
temperatures considered, the difference between the two dispersion
curves is at most of order of $10^{-5}$ for the transverse case and
decreases further as $k$ increases.  For the longitudinal case, the
relative difference is somewhat larger, but still negligible.  These
approximations greatly accelerate computations without significant
loss of accuracy.

\subsection{Checks of Differential Emissivities}
\label{sec:CHECKS}

The expressions for the differential emissivities in Sec. \ref{sec:DIFFEMI} can be integrated over
$(E, \theta)$ or $(E_1, E_2)$, in order to recover
results for the total emissivities calculated independently from
Eqs.~(\ref{QTTOTAL}) and (\ref{QLTOTAL}), respectively.  This is not
only a useful check of results derived for the differential
emissivities, but also allows for a qualitative understanding of photon and
plasmon decay under various physical conditions of the plasma.  In
this section, we analyze the transverse emissivity in some detail.
The longitudinal and axial emissivities are discussed in
Sec. \ref{sec:COMP}.   
\begin{figure}[h!]
\includegraphics[width=0.45\textwidth]{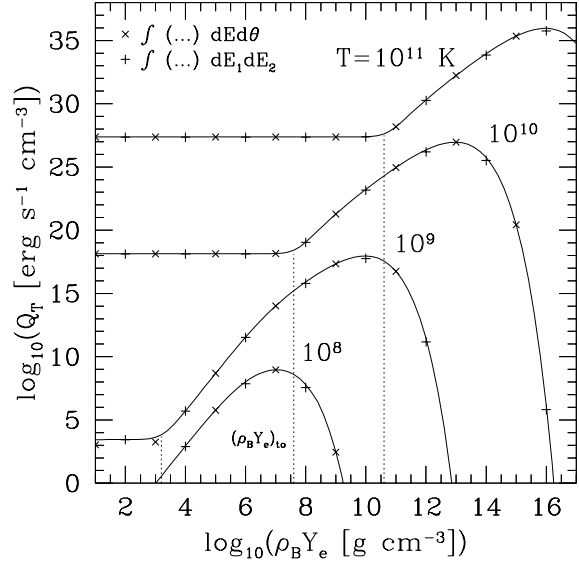} 
\caption{Total transverse emissivities $Q_T$ from Eqs.~(\ref{QTTOTAL})
versus baryon density at the indicated temperatures. The symbols
`$\times$' and `$+$' show results of
$\int\frac{d^2Q_T}{dEd\cos\theta}dEd\cos\theta$ and
$\int\frac{d^2Q_T}{dE_1dE_2}dE_1dE_2$ using the differential
emissivities from Eqs.~(\ref{dQET}) and (\ref{dQEE}), respectively. 
The vertical dotted lines mark the turn-on densities at which $Q_T$ 
abruptly changes its density dependence.}
\label{QTFIGURE}
\label{QLFIGURE}
\end{figure}

Integration over energies and angles were performed by substituting
the approximate expressions for $\Pi_T$, $Z_T$, and $J_T(E, \theta)$
and $J_T(E_1, E_2)$ in Eqs.~(\ref{ZTA}), (\ref{PITA}), (\ref{JETA}),
and (\ref{JEEA}).  The results are shown in Fig. \ref{QTFIGURE}.  It
is gratifying that there is excellent agreement between the
emissivities calculated in two different ways. In addition, they also
agree with those published in the literature \cite{BRAATEN1,BEAUDET2,
SCHINDER1}.

While the transverse total emissivities shown in
Fig.  \ref{QTFIGURE} increase rapidly  with temperature, their
behavior as a function of density is more complex.  At a
fixed temperature, the basic characteristics to note are: \\
(1) $Q_T$ is independent of the density $\rho_BY_e$  until a
turn-on density $(\rho_BY_e)_{\rm to}$ is reached, \\
(2) For densities larger than this turn-on density, $Q_T$ exhibits a
power-law rise until a maximum is reached at $(\rho_BY_e)_{\rm peak}$, and \\
(3) For $\rho_BY_e \gg (\rho_BY_e)_{\rm peak}$, the fall-off with density
is exponential.

In the next section, we proffer both qualitative and quantitative 
analyses of these features. 

\subsection{Qualitative Behaviors}
\label{sec:QUAL}

\begin{figure}[ht]
\begin{picture}(220, 440)(0, 0)
\put(0, 182){\includegraphics[width=0.45\textwidth]{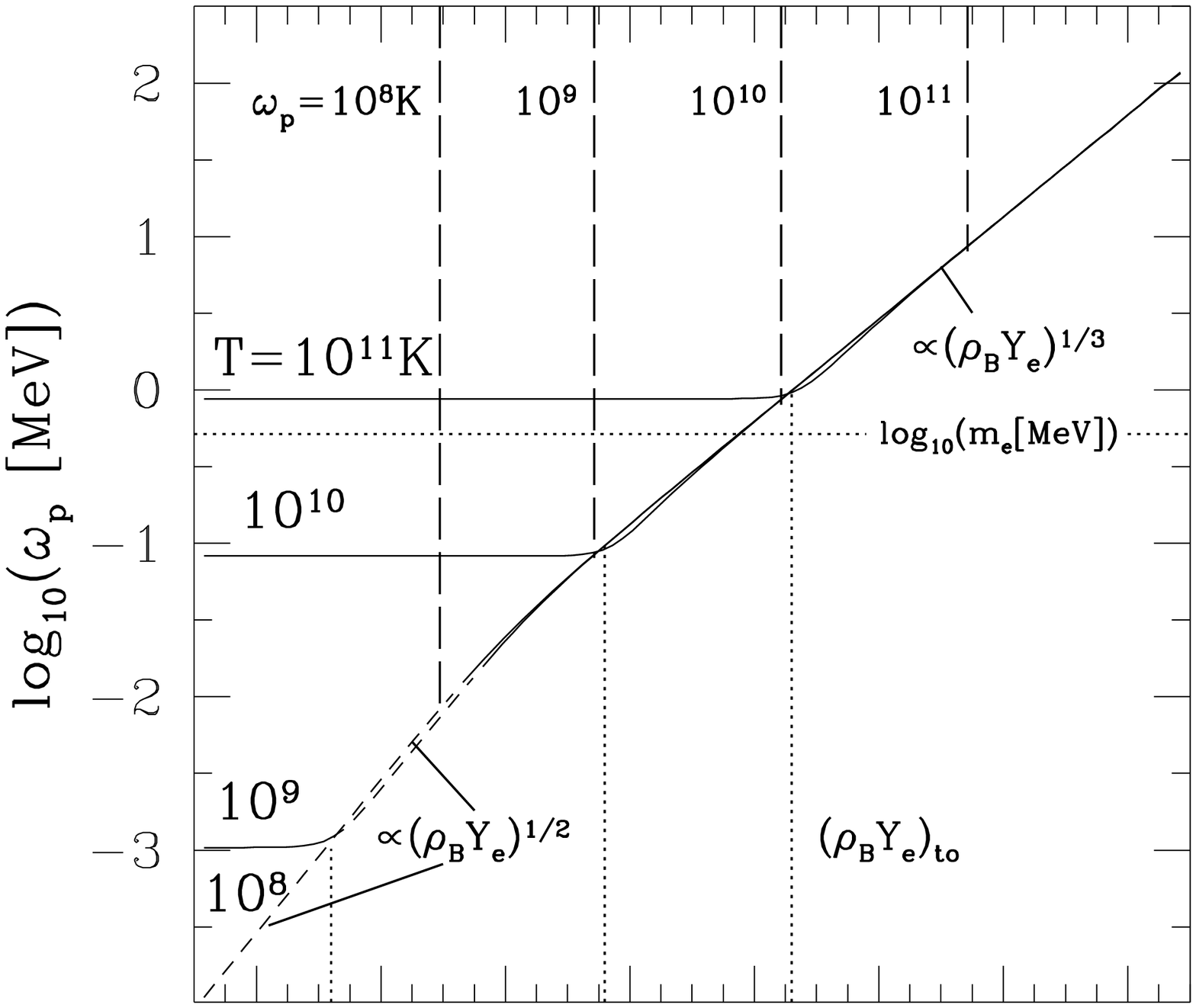}}
\put(180, 282){(a)}
\put(0, 0){\includegraphics[width=0.45\textwidth]{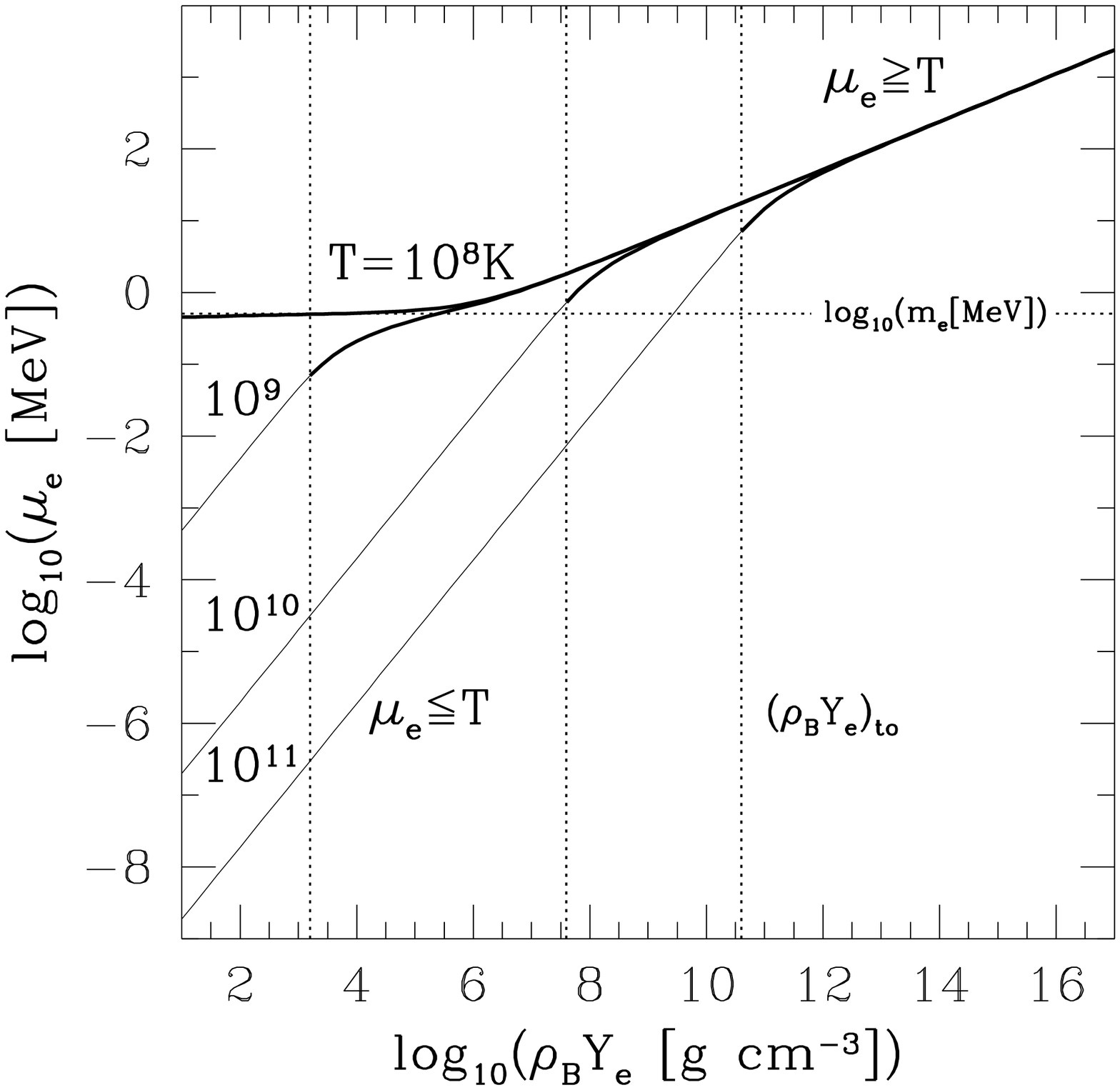}}
\put(180, 120){(b)}
\end{picture}
\caption{The plasma frequency $\omega_p$ (upper panel) and the
chemical potential $\mu_e$ (lower panel) as functions of density for
select temperatures.  The vertical dashed lines in panel (a) indicate
the densities at which $\omega_p = T$.  The vertical dotted lines mark
the turn-on densities at which the plasma frequency and the neutrino
emissivity (Fig.~\ref{QTFIGURE}) abruptly change their behavior from
being independent of density. }
\label{WPMUFIGURE}
\end{figure}

In order to gain a qualitative understanding of the basic features of
$Q_T$ in terms of the intrinsic properties of the plasma, it is
instructive to inspect the behaviors of the chemical potential $\mu_e$
(together with $T$, this determines $n_e$) and plasma frequency
$\omega_p$ (this is the characteristic energy scale generated by
interactions in the medium) as $T$ and $\rho_BY_e$ are varied.
Fig. \ref{WPMUFIGURE} shows $\mu_e$ (lower panel) obtained by inverting
Eq.~({\ref{MU}) and $\omega_p$ (upper panel) calculated from
Eq.~(\ref{WP}). The results in Fig. \ref{QTFIGURE} are readily
interpreted on the basis of the trends observed in
Fig. \ref{WPMUFIGURE}. To establish the main points, we focus on the
transverse emissivity $Q_T$ in this section.  The sub-leading
longitudinal and axial contributions can be understood in a similar
fashion (see Sec. \ref{sec:COMP}).

Noteworthy features of the plasma frequency in panel (a) of 
Fig.~\ref{WPMUFIGURE} are: \\
(1) $\omega_p$ is independent of $\rho_BY_e$ till the turn-on density
$(\rho_BY_e)_{\rm to}$ is reached (this is at the root of why $Q_T$ is
constant for $\rho_BY_e < (\rho_BY_e)_{\rm to}$), and \\
(2) $\omega_p$ shows a power-law increase for $\rho_BY_e >
(\rho_BY_e)_{\rm to}$, the index depending both on the extent to which
the plasma is in the non-degenerate, partially degenerate or
degenerate regime and on whether electrons are relativistic or
nonrelativistic.
 
The bold and light portions of the various curves in panel (b) of
Fig.~\ref{WPMUFIGURE} mark the regions of densities for which $\mu_e
\geq T$ and $\mu_e \leq T)$, respectively.  Inasmuch as $\mu_e \simeq T$
indicates partial degeneracy of the plasma, the bold and light
portions refer to the degenerate and non-degenerate conditions,
respectively.  For reference, the electron mass, which when compared
with $\mu_e$ or $T$ determines the degree of relativity, is marked by
the horizontal dotted line in this figure. The vertical dotted lines
show the respective locations of the turn-on densities.

\begin{figure}[ht]
\includegraphics[width=0.45\textwidth]{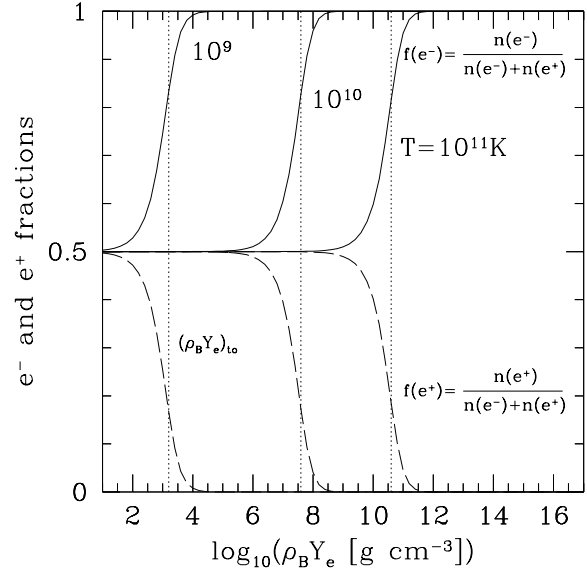} 
\caption{Electron and positron fractions in the plasma versus density
for select temperatures. The vertical dotted lines mark the densities
at which the plasma frequency (Fig.~\ref{WPMUFIGURE}) and the neutrino
emissivity (Fig.~\ref{QTFIGURE}) abruptly change their density 
dependence. }
\label{fracs}
\end{figure}

In order to assess the role of positrons in determining the behavior
of the plasma frequency, we show the electron and positron fractions
versus density in Fig.~\ref{fracs}.  For a given temperature, the
positron fraction begins to decrease with increasing density or
chemical potential and vanishes exponentially in the limit of complete
degeneracy. It is interesting that the positron fraction is of order
15\% at the turn-on densities for each temperature shown in
Fig.~\ref{fracs}. Although this is small compared to the electron
fractions at these densities, it will be necessary to account for the
presence of positrons to gain a quantitative understanding of
$\omega_p$ as a function of density and temperature. Since varying
degrees of degeneracy and relativity are encountered in the wide
ranges of density and temperature considered here, we consider below
those limiting cases that help us to understand the main features of
$\omega_p$ and $Q_T$.

\subsubsection{Plasma Frequency}
\label{sec:PF}

We begin by enquiring why $\omega_p$ is constant with increasing
$\rho_BY_e$ at low densities.  Consider first the cases of
$T=10^{11}$K and $10^{10}$K, both of which exceed the electron mass
($m_e = 5.92\times 10^9~{\rm K})$.  The plasma is thus
in the relativistic regime in which the net electron density and
the plasma frequency are well approximated by
\begin{eqnarray}
\label{nmulT1011}
n_e &=& \frac{\mu_e}{3\pi^2}\biggl(
\mu_e^2+T^2\pi^2\biggr) \simeq \frac{\mu_e T^2}{3} \,, \nonumber \\
\omega_p^2 &=& \frac{4\alpha}{3\pi}\biggl(\mu_e^2+
\frac{\pi^2T^2}{3}\biggr)\simeq \frac{4\pi \alpha T^2}{9} \,.
\label{wpmulT1011}
\end{eqnarray}
In writing the rightmost relations above, we have used the fact that
$\mu_e \ll T$ ({\it i.e.,} the plasma is non-degenerate) as is evident
from the lower two curves in panel (b) of Fig. \ref{WPMUFIGURE}.  The
plasma frequency is therefore effectively independent of
$\rho_BY_e$ and is given by 
\begin{equation}
(\omega_p)_{\rm {pl}} \simeq {T}/{10} \,,
\end{equation}
where the subscript ``pl'' denotes the plateau value.  With
increasing density, the chemical potential $\mu_e$ rises linearly with
density (see Eq.~(\ref{nmulT1011})) until the plasma enters the
partially degenerate regime in which $\mu_e$ becomes comparable to
$T$. In this regime, $\omega_p$ abruptly changes its behavior from a
constant to a power-law behavior. An estimate of the density,
$(\rho_BY_e)_{\rm to}$, at which this turn on occurs can be obtained
by setting $\mu_e \simeq T$ in Eq.~(\ref{nmulT1011}), whence we have
\begin{eqnarray}
\label{To1011}
\begin{array}{rcl}
{\displaystyle (n_e)_{\rm to} }& \simeq &{\displaystyle  \frac{T^3}{2.72} } 
\quad {\rm or}  \\  
{\displaystyle \left({\rho_B}{Y_e}\right)_{\rm to} }
& \simeq & {\displaystyle8 \times 10^7 \left(\frac{T}{{\rm MeV}}\right)^3}~
{\rm g~cm^{-3}} \,.
\end{array}
\Bigg\} \quad {\rm Relativistic} 
\end{eqnarray}
For $T=10^{11}$K (8.62 MeV) and  $10^{10}$K (0.862 MeV),
Eq.~(\ref{To1011}) yields  $\log_{10}~(\rho_BY_e)_{\rm to} = 10.7$ and
7.7, respectively, in agreement with the results shown in  
Fig. \ref{WPMUFIGURE}.  

Further increase in density results in the plasma becoming degenerate
as $\mu_e \gg T$.  Using the Sommerfeld expansion, we find the net
electron density and plasma frequency in this region to be
\begin{eqnarray}
\label{nmuhT1011}
n_e &=& \frac{1}{3\pi^2}\biggl(
{p_F^3}+\frac{ \pi^2T^2}{2p_F}{(2p_F^2+m_e^2)}\biggr) \simeq
\frac{p_F^3}{3\pi^2} \,, \\
\omega_p^2 &=& \frac{4\alpha}{3\pi}\biggl(p_F^2v_F+
\frac{\pi^2T^2}{6E_F^2}\biggl[\frac{3m_e^2}{v_F}
+2p_F^2v_F\biggr]\biggr) \simeq \frac{4\alpha}{3\pi}p_F^2v_F \,,
\label{wpmuhT1011}
\end{eqnarray}
where $E_F =(p_F^2+m_e^2)^{1/2}$ and $v_F=p_F/E_F$. The rightmost
relations here refer to the case in which the thermal contributions
are small, {\it i.e.}, $T \ll \mu_e \simeq E_F$.  In this regime, the
plasma frequency varies with the net electron density as $\omega_p
\propto n_e^{1/3}$, since $v_F \rightarrow 1$.

For $T=10^{9}$K, the region of densities in which $\omega_p$ is
constant lies in the classical regime, in which $T \ll m_e$ and
$m_e-\mu_e \gg T$. Including the contributions from positrons (this was
neglected in Ref.~\cite{BRAATEN2}), we find that the net electron
density and the plasma frequency in this non-degenerate and
non-relativistic limit are
\begin{eqnarray}
\label{nmulT9}
n_e &=& 4 \sinh\left(\frac{\mu_e}{T}\right) 
\left(\frac{m_eT}{2\pi}\right)^{3/2} e^{-m_e/T}   
\biggl(1+\frac{15T}{8m_e}\biggr) \,, \\
\left(\omega_p^2\right)_{\rm pl} &=& 
\frac {16\pi\alpha}{m_e} \cosh\left(\frac{\mu_e}{T}\right) 
\left(\frac{m_eT}{2\pi}\right)^{3/2} e^{-m_e/T} 
\biggl(1-\frac{5T}{8m_e}\biggr) \nonumber \\ 
& = &  \frac{4\pi\alpha n_e}{m_e} \coth\left(\frac{\mu_e}{T}\right)  
\biggl( 1-\frac{5T}{2m_e}\biggr) \,.
\label{wpmulT9}
\end{eqnarray}
For low densities at $T=10^9$K, $\mu_e \ll T$ which renders $\omega_p$
to be independent of $\mu_e$ and to be a function of $T$ alone and thus
independent of $n_e$.  At this temperature, Eq.~(\ref{wpmulT9}) yields
$\log_{10}~\omega_p \simeq -2.95$, in close agreement with the exact result
shown in Fig.~\ref{WPMUFIGURE}.  The turn-on density at which
$\omega_p$ begins its characteristic power-law rise with density can be
obtained by setting $\mu_e\simeq T$ in Eq.~(\ref{nmulT9}).  At this
temperature, we find $\log_{10}~(\rho_BY_e)_{\rm to} \simeq 3.2$, in
excellent agreement with the exact results shown in Fig.
\ref{WPMUFIGURE}.

For $\rho_BY_e > (\rho_BY_e)_{\rm to}$, we observe that $\omega_p\propto
(\rho_BY_e)^{1/2}$ as the plasma begins to enter the partially
degenerate regime from the non-degenerate regime.  For densities
characterized by $\mu_e \gg T$, the plasma frequency exhibits the same
degenerate behaviour as for the higher temperature cases, {\it i.e.,}
$\omega_p\propto (\rho_BY_e)^{1/3}$. 

For $T=10^8$K, the region of densities in which $\omega_p$ is constant
with density lies well below the lowest density shown in Fig.
\ref{WPMUFIGURE}.  Note that the $(\rho_BY_e)^{1/2}$ rise, typical of
a non-degenerate plasma, is also observed for a degenerate ($\mu_e >>
T$), but semi-relativistic ($m_e \sim T$) plasma. The feature that
$\mu_e$ is nearly constant until a density of $10^6~{\rm g~cm^{-3}}$ is
connected with the fact that in this highly degenerate case,
\begin{eqnarray}
\mu_e = \left[m_e^2 + (3\pi^2n_e)^{2/3} \right]^{1/2} \,
\end{eqnarray} 
begins to rise significantly with density only for densities that
exceed the threshold density of $m_e^3/(3\pi^2)$. This situation does
not occur for the other temperatures shown in Fig.  \ref{WPMUFIGURE}
to the degree that it does for $T=10^8$K.

We turn now to determine the density at which $\omega_p = T$.
Consider first the degenerate case in which $\mu_e \gg T$ and the plasma
is relativistic ($v_F \rightarrow 1$ and $\mu_e \rightarrow
p_F$). Ignoring the small temperature dependent corrections,
Eqs.~(\ref{nmuhT1011}) and (\ref{wpmuhT1011}) can be combined to yield
\begin{eqnarray}
\begin{array}{rcl}
{\displaystyle\left(\rho_BY_e\right)_{\omega_p=T}} &=& {\displaystyle 
\frac {m_p}{3\pi^2} \left(\frac {3\pi}{4\alpha} \right)^{3/2} T^3 \,,}
\\  
&\simeq& {\displaystyle 4.26 \times 10^{10} \left(\frac {T}{\rm MeV}\right)^3~ 
{\rm g~cm^{-3}} \,.}
\end{array}
\Bigg\}\quad {\rm Relativistic}     
\end{eqnarray}
Results from this expression agree closely with the numerically
calculated exact results (see the vertical dashed lines in panel (a)
of Fig.
\ref{WPMUFIGURE}) for all but the lowest temperature of $T=10^8$K
for which Eqs.~(\ref{nmuhT1011})
and (\ref{wpmuhT1011}) can still be used to advantage, but with
$v_F=p_F/E_F$.  The leading order solution of the resulting cubic
equation in $p_F^2$ leads to
\begin{eqnarray}
\begin{array}{rcl}
{\displaystyle\left(\rho_BY_e\right)_{\omega_p=T}} &=&  
{\displaystyle\frac{m_pm_eT^2}{4\pi\alpha}} \\ 
&\simeq& {\displaystyle1.22 \times 10^9  \left(\frac {T}{\rm MeV}\right)^2~ 
{\rm g~cm^{-3}} \,, } 
\end{array}
\Bigg\}\quad {\rm Nonrelativistic}     
\end{eqnarray}
which agrees with the exact result at  $T=10^8$K.

\subsubsection{Transverse Emissivity}
\label{sec:TE}

The results of the previous section enable us to organize the
discussion of the basic qualitative features of the transverse
emissivity $Q_T$, which is largely governed by the temperature $T$ and
the plasma frequency $\omega_p(\mu_e,T)$.  Depending on the extent to
which 
the plasma is degenerate and relativistic, $\omega_p$ is
either a simple function of $\mu_e$ or $T$ alone, or an involved
combination of both $\mu_e$ and $T$.  The density regions where
$\omega_p \leq T$ and $\omega_p \geq T$ identified in the previous
section are particularly useful here as they enable us to examine the
limiting cases $\omega_p \ll T$ and $\omega_p \gg T$ in which $Q_T$
in Eq.~(\ref{QTTOTAL}) can be cast in physically transparent forms
\cite{BRAATEN1}.

In the case of $T \gg \omega_p$, the main contribution to 
$Q_T$ comes from high photon momenta.  To a very good
approximation \cite{BRAATEN1},
\begin{equation}\label{QTAT}
Q_T \simeq \frac{2\sum_f {(C_V^f)}^2 
G_F^2}{48 \pi^4 \alpha} \zeta(3) m_T^6 T^3
\,,
\end{equation}
where $\zeta(3) \simeq 1.202$ is Riemann's Zeta function and 
$m_T$ is the transverse photon mass (this can be read off from the
high energy limit of the transverse
polarization function, {\it i,e.,} $\Pi_T(\omega_T, k)
\xrightarrow{k\rightarrow \infty} m_T^2$; the transverse mass lies in
the range $\omega_p \leq m_T \leq {\sqrt {3/2}}~\omega_p )$.   
With $m_T = \omega_p$,  
\begin{equation}\label{NQTAT}
(Q_T)_{\rm pl} 
\simeq 2.8\times 10^{24}~ \omega_p^6 ~T^3~~{\rm erg~s^{-1}~cm^{-3}}
\,,
\end{equation}
where $\omega_p$ and $T$ are in units of MeV.  As discussed in the
above section, $\omega_p$ is independent of density for $\rho_BY_e
\leq (\rho_BY_e)_{\rm to}$. Consequently, $Q_T$ is also density
independent up to $(\rho_BY_e)_{\rm to}$, a trend which is maintained
at all temperatures whenever $\omega_p$ is density independent.  Use
of Eq.~(\ref{NQTAT}) with $\omega_p$ from Eqs.~(\ref{wpmulT1011}) and
(\ref{wpmulT9}) yields excellent agreement with the plateau results at all
temperatures shown in Fig.~\ref{QTFIGURE}.

For $T \ll \omega_p$, $Q_T$ takes the form  
\cite{BRAATEN1} 
\begin{equation}\label{QTAW}
        Q_T \simeq \frac{\sum_f {(C_V^f)}^2 G_F^2}{48 \pi^4 \alpha} 
\sqrt{\frac{\pi}{2}}  \omega_p^{15/2} T^{3/2} e^{-\omega_p/T} \,.
\end{equation}
This expression allows us to qualitatively understand the power-law
rise of $Q_T$ for densities $\rho_BY_e \geq (\rho_BY_e)_{\rm to}$.
In the classical or non-degenerate regime, $\omega_p\propto
n_e^{1/2}$, whereas in the degenerate regime, $\omega_p\propto
n_e^{1/3}$. As a result, 
\begin{equation}
Q_T \propto (\rho_BY_e)^p\,, \qquad  2.5 \leq
p \leq 3.75.
\end{equation}
For all temperatures considered, the peak value of $Q_T$ occurs in the
relativistic and degenerate regime in which $T \ll \omega_p$.  Noting
that at the peak $\omega_p = 7.5~T$, we can utilize Eq.~(\ref{QTAW})
to obtain
\begin{equation}\label{NQTAW}
(Q_T)_{\rm peak} \simeq 2.9 \times 10^{27}~
\left(\frac{T}{\rm MeV}\right)^9~~{\rm erg~s^{-1}~cm^{-3}} \,, 
\end{equation}
which yields the peak values in Fig.~\ref{QTFIGURE}. The density at
which the peak occurs may be obtained by inverting  
Eq.~(\ref{wpmuhT1011}) with the result  
\begin{equation}
(n_e)_{\rm peak}  \simeq \frac {1}{3\pi^2} 
\left(\frac {3\pi}{4\alpha}\right)^{3/2}~\omega_p^3 \,.      
\end{equation}
Using the fact that $\omega_p = 7.5~T$ at the peak, we arrive at 
\begin{eqnarray}
\left(\rho_BY_e\right)_{\rm peak} &=& \frac{3375}{32\alpha}~
\left(\frac{3}{4\pi\alpha} \right)^{1/2}~
m_p\left(\frac{T}{\rm MeV}\right)^3  \nonumber \\
&\simeq& 1.8 \times 10^{13}~\left(\frac{T}{\rm MeV}\right)^3~
{\rm g~cm^{-3}} \,,  
\end{eqnarray}
which accounts for the results at all temperatures shown in 
Fig.~\ref{QTFIGURE}.

As the net electron density increases to values well in excess of
$(\rho_BY_e)_{\rm peak}$, the plasma frequency begins to become
significantly larger than the temperature.  In this case, the
emissivity is exponentially damped by the Boltzmann factor
$e^{-\omega_p/T}$ as seen in Fig. \ref{QTFIGURE}.

\subsubsection{Comparison of Transverse, Longitudinal, and Axial Emissivities}
\label{sec:COMP}

The qualitative analysis of the longitudinal and axial 
emissivities can be carried out along the same lines as that for
the transverse emissivity.  Figure \ref{QTLA} shows the individual
contributions from the three channels at $T=10^{11}$K and $T=10^9$K,
respectively.  The symbols ``$\times$'' and ``$+$'' show results
obtained by integrations of the differential emissivities in Sec. \ref
{sec:DIFFEMI}.

The plateau value of $Q_L$ can be inferred by considering the limit 
$T \gg \omega_p$. In this case,  
\begin{equation}
Q_L \simeq  \frac{\sum_f {(C_V^f)}^2 G_F^2}{96 \pi^4 \alpha}
A(v_*)~\omega_p^8~T \,,
\label{QLPLA} 
\end{equation}
which shows that $Q_L$ is suppressed relative to $Q_T$ by a factor
$\omega_p^2/T^2$ \cite {BRAATEN1}.  In the relativistic limit, the
coefficient $A(v_* \rightarrow 1) \simeq 0.349$~\cite{BRAATEN1}.
Using the result $\omega_p^2 = 4\pi\alpha T^2/9$ from
Eq.~(\ref{wpmulT1011}) that is valid for $\mu_e \ll T$, we find
\begin{equation}
(Q_L)_{\rm pl} \simeq 2.14 \times 10^{15}~ 
\left(\frac{T}{{\rm MeV}}\right)^9~{\rm erg~cm^{-3}~s^{-1}} \,,
\end{equation}
which is in fair agreement with the exact result for $T=10^{11}$K shown in
Fig.~\ref{QTLA}.  

The turn-on density and power-law rise of $Q_L$ with density are
similar those of $Q_T$.  As for $Q_T$, the peak value of $Q_L$ occurs
in the regime $T \ll \omega_p$. In this case, $Q_L$ has nearly the
same functional form as $Q_T$ in Eq.~(\ref{QTAW}), but with a slightly
different numerical coefficient (for $v_F \rightarrow 1$, $Q_L \simeq
{\sqrt 2}~Q_T$) \cite{BRAATEN1}. For a fixed temperature, therefore,
the peaks of $Q_L$ and $Q_T$ occur at the same density with nearly the
same values.

We discuss now the relatively small contribution of the axial
channel and focus on its peak value, since the plateau region for this
case is at densities substantially lower than those of interest here.
As shown in Ref.~\cite{BRAATEN1},
\begin{eqnarray}
Q_A &\simeq& 
\frac{\sum_f {(C_A^f)}^2 G_F^2}{96 \pi^4 \alpha} 
3\sqrt{2\pi} 
~\left( 1 + \frac 15 v_*^2 \right)^{-5/2}
\omega_p^{9/2} \omega_A^2 T^{5/2} e^{-\omega_p/T} \,
\end{eqnarray} 
for $T \ll \omega_p$.  In the degenerate and relativistic limit, ($T
\ll \mu_e$ and $v_* \rightarrow v_F \rightarrow 1$), it is
straightforward to establish that the factor $\omega_A \simeq (\alpha/
3\pi)^{1/2}\omega_p$.  This implies that the peak of $Q_A$ occurs at 
$\omega_p = 6.5~T$. Exploiting this result, we arrive at 
\begin{equation}
(Q_A)_{\rm peak} \simeq 5 \times 10^{23} 
\left(\frac{T}{{\rm MeV}}\right)^9~{\rm erg~cm^{-3}~s^{-1}} \,,  
\end{equation}
which nicely reproduces the exact results at $T=10^{11}$K and
$T=10^9$K shown in Fig.~\ref{QTLA}.  

Eqs. (\ref{nmuhT1011}) and (\ref{wpmuhT1011}), coupled with $\omega_p
= 6.5~T$, yield the density at which $Q_A$ peaks. Explicitly,  
\begin{equation}
\left(\rho_BY_e\right)_{\rm peak}
\simeq 1.17 \times 10^{13}   \left(\frac {T}{\rm MeV}\right)^9~ 
{\rm g~cm^{-3}} \,. 
\end{equation}
This density is slightly smaller than the density at which
$Q_L$ and $Q_T$ attain their peaks, since $\omega_p = 6.5~T$ at the
peak of $Q_A$, while $\omega_p = 7.5~T$ for the other two channels.

Our results for $Q_A$ differ by orders of magnitude with those in
Fig. 2 of Ref.~\cite{BRAATEN1}, although the formal expressions in
both works are identical. For example, at the maximum density of
~~$\log_{10} \rho_BY_e = 14$ at $T=10^{11}$K shown in
Ref.~\cite{BRAATEN1}, $\log_{10}~Q_A \simeq 32.8$ whereas we obtain
$\simeq 31$.  At the lowest density of~~ $\log_{10} \rho_BY_e \simeq
10.5$ displayed there, $\log_{10}~Q_A \simeq 24$, while we find $\simeq
23.4$. This latter result can also be verified from the analytic
expression in Eq.~(34) of Ref.~\cite{BRAATEN1}, which is valid in the
regime $T \gg \omega_p$.  We have been unable to resolve these large
discrepencies in the numerical results for $Q_A$. Since we agree with
all of the formal results derived in Ref. \cite{BRAATEN1}, we
attribute these differences to a possible numerical error in the
calculations of $Q_A$ in Ref. \cite{BRAATEN1}.  We wish to emphasize,
however, that these discrepencies are not of much significance, since
$Q_A$ is orders of magnitude lower than both $Q_L$ and $Q_T$ for the
densities and temperatures of interest here.

\begin{figure}[ht]
\includegraphics[width=0.45\textwidth]{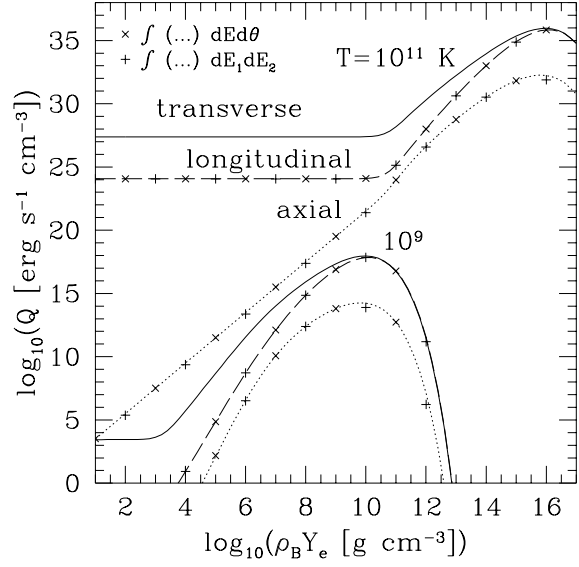}
\caption{Individual contributions from the transverse,
longitudinal, and axial channels to the neutrino emissivity. }
\label{QTLA}
\end{figure}

\subsection{Features of Differential Emissivities}
\label{sec:SPECIAL}

An interesting dependence of the differential emissivity on the angle
between the two outgoing particles is worth noting.  If, for example,
the differential emissivity from Eq. (\ref{dQET}) is integrated over 
the outgoing energy,
the only remaining dependence is on the
outgoing angle $\theta$.  For a massive particle decaying into two
particles, we would expect ``back-to-back'' decay, {\it i.e.}, the two
outgoing particles prefer to be emitted at $\theta=180^\circ$, for
which ${dQ_T/d(\cos\theta)}$ should exhibit a maximum.  The
massiveness of the photon can be gauged by the photon effective mass
which strongly depends on the properties of the medium, such as its
temperature, chemical potential, and net electron
density. Effectively, the photon mass is roughly equal to $\omega_p$
and its kinetic energy is of order $T$. For $\omega_p$ much larger
than the kinetic energy, we can expect aspects associated with the
decay of a massive particle.  However, if its mass is negligible
compared to the kinetic energy, we should see a significantly
different behavior.

\begin{figure}[ht]
\begin{picture}(220, 220)(0, 0)
\put(0, 0){\includegraphics[width=0.45\textwidth]{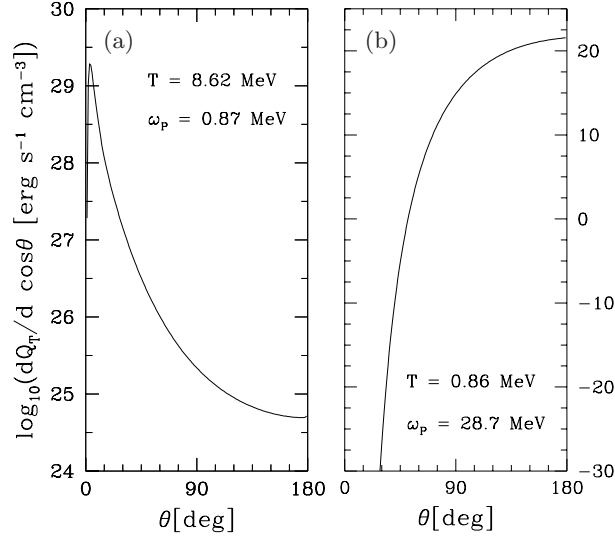}}
\put(35, 195){(a)}
\put(134, 195){(b)}
\end{picture}
\caption{Angular dependence of the differential emissivity for two
extreme cases. Panel (a) shows the dominance of energy emitted at a
small outgoing angle between the neutrinos for densities in the
plateau range $\rho_BY_e=10-10^{10}\,{\rm g/cm}^3$, while panel (b) depicts a
situation in which ``back to back'' emission is preferred for
densities in the peak range $\rho_BY_e=10^{12}-10^{13}\,{\rm g/cm}^3$.}
\label{INTEG}
\end{figure}

These different behaviors are illustrated in the results presented in
Fig. \ref{INTEG}.  In panel (a), the ``light'' photon
(i.e. $\omega_p\ll T$) decay is shown. In this case, the main
contribution to the total emissivity comes from neutrino pairs with
small outgoing angles between them. This corresponds to a high Lorentz
boost of the photon with respect to the rest frame where calculations
were performed and the angle that was $\approx 180^\circ$ in the
photon rest frame gets significantly shifted.  When the photon effective
mass becomes the dominant energy scale ($\omega_p\gg T$), the result is
the massive particle decay depicted in panel (b).


\section{Kernels for Neutrino Transport Calculations }
\label{sec:PKER}

In this section, we discuss briefly how the differential emissivities
calculated in this work enter in calculations of neutrino transport.
The evolution of the neutrino distribution function $f$ is generally
described by the Boltzmann transport equation in conjunction with
hydrodynamical equations of motion together with baryon and lepton
number conservation equations. Ignoring general  relativistic effects for
simplicity (see, for example, Ref. \cite {BRUENN1} for full details),
the Boltzmann equation is  
\begin{eqnarray}
\label{BOLTZMANN}
\frac{\partial f}{\partial t}+v^i\frac{\partial f}{\partial x^i}+
\frac{\partial (f F^i)}{\partial p^i}
=B_{EA}(f)+B_{NES}(f)+B_{\nu{\cal N}}(f)+B_{TP}(f) \,,
\end{eqnarray}
where $F^i$ is the force acting on the particle. 
In the source term on the right hand side, 
$B_{EA}(f)$ incorporates neutrino emission and
absorption processes, $B_{NES}(f)$ accounts for the neutrino-electron
scattering process, $B_{\nu{\cal N}}(f)$ includes scattering of
neutrinos off nucleons and nuclei, and $B_{TP}(f)$ considers
the thermal production and absorption of neutrino-antineutrino pairs.

In discussing the contribution of the plasma process to $B_{TP}$, we
follow Ref. \cite{BRUENN1} in which neutrino pair production from
$e^+e^-$ annihilation was considered in detail.
Suppressing the dependencies on $(r,t)$ for notational simplicity, 
the source term for the plasma process can be written as
\begin{widetext}
\begin{eqnarray}\label{BTP}\nonumber
B\big(f(\mu_1,E_1)\big) &=& 
\biggl[1-f(\mu_1,E_1)\biggr]\frac{1}{(2\pi)^3}
\int_0^\infty E^2_2 \,\,dE_2 \int_{-1}^1 
d\mu_2 \int_0^{2\pi}d\phi_2 \,R^p
\biggl(E_1,E_2,cos\,\theta\biggr)
\,\biggl[1-\bar{f}(\mu_2,E_2)\biggr]\\
&-&  f(\mu_1,E_1)\frac{1}{(2\pi)^3}
\int_0^\infty E^2_2 \,\,dE_2 \int_{-1}^1 d\mu_2 
\int_0^{2\pi}d\phi_2
\,R^a\biggl(E_1,E_2,cos\,\theta\biggr)
\,\bar{f}(\mu_2,E_2) \,,
\end{eqnarray}
\end{widetext}
where the first and the second terms correspond to the source
(neutrino gain) and sink (neutrino loss) terms, respectively. Angular
variables $\mu_i\equiv \cos\theta_i$ and $\phi_i$ are defined with
respect to the $z$-axis that is locally set parallel to the outgoing
radial vector ${\bf r}$. The angle $\theta$ between the
neutrino and antineutrino pair is related to $\theta_1$ and $\theta_2$
through 
\begin{eqnarray}
\cos\theta &=&
\mu_1\mu_2+\sqrt{(1-\mu_1^2)(1-\mu_2^2)}~\cos(\phi_1-\phi_2) \,.
\end{eqnarray}
The production and absorption kernels are given by
\begin{eqnarray}\label{kernel}
R^{\genfrac{}{}{0pt}{}{p}{a}}(E_1, E_2, \cos\theta)&=& 
\int \frac{d^3k}{(2\pi)^3}\, Z_Y(k)\,
\genfrac{(}{)}{0pt}{}{\xi n_B(\omega, T)}{1-n_B(\omega, T)} 
\frac{1}{8\omega E_1 E_2}  \,
\delta^4(K-Q_1-Q_2)(2\pi)^4\langle {|\mathcal{M}|}^2\rangle\,, 
\end{eqnarray}
where the subscript $Y$ stands for $T$--``transverse'' or
$L$--``longitudinal''. The factor $\xi$ accounts for the spin summation; $\xi=2$ for the transverse, axial and mixed cases,
while for the longidudinal case $\xi=1$.

The angular dependences in the kernels $R^{\genfrac{}{}{0pt}{}{p}{a}}(E_1, E_2,
\cos\theta)$ are often expressed in terms of Legendre polynomials as
\begin{eqnarray}\label{RLegendre}
R^{\genfrac{}{}{0pt}{}{p}{a}}(E_1, E_2, \cos\theta)&=&
\sum_{l=0}^{\infty} \frac{2l+1}{2} \Phi^{\genfrac{}{}{0pt}{}{p}{a}}_l(E_1, E_2) 
P_l(\cos\theta) \,, 
\end{eqnarray}
where the Legendre coefficients
$\Phi^{\genfrac{}{}{0pt}{}{p}{a}}_l(E_1, E_2)$ depend exclusively on
energies.
 
From Eq. (\ref{kernel}), it is evident that the kernels are related
to the neutrino rates and emissivities.  We first consider the
production kernel $R^{p}(E_1, E_2, \cos\theta)$.  The corresponding
analysis for the absorption kernel $R^{a}(E_1, E_2, \cos\theta)$ can
be made along the same lines, but with the difference that $n_B$ is
replaced by $1-n_B$.  The neutrino production rate is given by
\begin{eqnarray}\label{rate}
\Gamma&=&\xi\,\int \frac{d^3k}{(2\pi)^3 2\omega}\, Z_Y(k)\, \frac{d^3q_1}{(2\pi)^3 2E_1}
\frac{d^3q_2}{(2\pi)^3 2E_2}
 \, n_B(\omega,T) \, \delta^4(K-Q_1-Q_2)(2\pi)^4\langle{|\mathcal{M}|}^2\rangle \nonumber \\
&=& \int \frac{d^3q_1}{(2\pi)^3} \frac{d^3q_2}{(2\pi)^3} 
R^{p}(E_1, E_2,\cos\theta)\,, 
\end{eqnarray}
which defines the kernel $R^p(E_1, E_2,\cos\theta)$ and is to be
identified with that in Eq.~(\ref{kernel}).
The emissivity $Q$ can also be cast in terms of $R^p$ using 
\begin{equation}
Q=\int \frac{d^3q_1}{(2\pi)^3} \frac{d^3q_2}{(2\pi)^3} (E_1+E_2) R^{p}(E_1,
E_2, \cos\theta)\,. \label{Remiss}
\end{equation}
Equations (\ref{rate}) and (\ref{Remiss}) can be inverted to obtain
\begin{eqnarray}\label{RQ}
R^{p}(E_1, E_2, \cos\theta) & = &
\frac{8\pi^4}{E_1^2 E_2^2}~\frac{d^3 \Gamma}{dE_1 dE_2 d\cos\theta}\nonumber \\
&=&\frac{8\pi^4}{E_1^2 E_2^2 (E_1+E_2)}~\frac{d^3 Q}{dE_1 dE_2 d\cos\theta}\,.
\end{eqnarray}
Utilizing the results from Eqs. (\ref{dQEET}), (\ref{dQLEET}),
(\ref{dQAEET}), and (\ref{dQMEET}) in Eq. (\ref{RQ}), the production
kernels for the transverse, longitudinal, axial and mixed vector-axial
parts are:
\begin{widetext}  
\begin{eqnarray}\label{kernelT}
R^{p}_T(E_1, E_2, \cos\theta)&=&
        \frac{G_F^2{\sum_f (C_V^f)}^2}{2\alpha}
        \, Z_T(k) \, n_B(\omega_T,T) \, \frac{\Pi_T^2(\omega_T, k)}{\omega_T} 
        \frac{(E_1-E_2)^2+E_1 E_2 (1+ \cos \theta)}
        {E_1^2 + E_2^2 + 2E_1 E_2 \cos \theta}\nonumber\\&& \times \quad  
        (1- \cos \theta)\; \delta(\omega_T - E_1 - E_2) \,, \\
R^{p}_L(E_1, E_2, \cos\theta)&=&
        \frac{G_F^2{\sum_f (C_V^f)}^2}{2\alpha}
        \, Z_L(k) \, n_B (\omega_L,T) \, (\omega_L^2-{k}^2)^2  \frac{E_1 E_2}{\omega_L} 
        \frac{1-\cos^2\theta} {k^2}
        \delta(\omega_L- E_1 - E_2)
        \Theta(\mathcal{K})\,, \label{kernelL}\\
R^{p}_A(E_1, E_2, \cos\theta)&=&
        \frac{G_F^2{\sum_f (C_A^f)}^2}{2\alpha}
        \, Z_T(k) \, n_B(\omega_T,T) \, \frac{\Pi_A^2(\omega_T, k)}{\omega_T} 
        \frac{(E_1-E_2)^2+E_1 E_2 (1+ \cos \theta)}
        {E_1^2 + E_2^2 + 2E_1 E_2 \cos \theta}\nonumber\\
&& \times \quad(1- \cos \theta)\,\delta(\omega_T- E_1 - E_2)\,,\label{kernelA} \\
R^{p}_M(E_1, E_2, \cos\theta)&=&
        \frac{G_F^2{\sum_f C_A^f C_V^f}}{\alpha}
        \, Z_T(k) \, n_B(\omega_T,T) \, 
        \frac{\Pi_A(\omega_T, k) \Pi_T(\omega_T, k)}{\omega_T k} (E_2-E_1)
        \nonumber \\ && \times
	\quad (1- \cos \theta)\; \delta(\omega_T- E_1 - E_2)\,.
	\label{kernelM}  
\end{eqnarray}
\end{widetext}
A simplifying feature in these expressions is worth noting. Since the
two outgoing neutrinos stem from the decay of a single photon or
plasmon, an energy-conserving $\delta$--function survives in the
expressions for the kernels. This is fortunate, since it makes the
calculation of the Legendre coefficients
$\Phi^{\genfrac{}{}{0pt}{}{p}{a}}_l(E_1, E_2)$ straightforward.  
The Legendre coefficients are determined from
\begin{eqnarray}\label{Legendre}
\Phi^{\genfrac{}{}{0pt}{}{p}{a}}_l(E_1, E_2) 
&=&\int_{-1}^1 d(\cos\theta) P_l(\cos\theta) 
R^{\genfrac{}{}{0pt}{}{p}{a}}(E_1, E_2, \cos\theta)\,.
\end{eqnarray}
Integrating over the $\delta$--function introduces a Jacobian and sets
$\cos\theta$ to be
\begin{eqnarray}
cos{\tilde \theta} = \frac{k^2-E_1^2-E_2^2}{2E_1E_2}\,.
\end{eqnarray}
As a result, the Legendre coefficients for the production kernels are:
\begin{eqnarray}\label{phiT}
\Phi^p_T(E_1, E_2)&=&\frac{G_F^2{\sum_f(C_V^f)}^2}{2\alpha}
        \, Z_T(k) \, n_B(\omega_T,T) \, \frac{\Pi_T^2(\omega_T, k)}{\omega_T E_1 E_2} 
        \bigg[E_1E_2 - \frac{(k^2+E_1^2-E_2^2)(k^2-E_1^2+E_2^2)}{4k^2} \bigg] 
        \nonumber \\
        && \times \quad  J_T(E_1, E_2) \, 
        \Theta(4E_1E_2-\Pi_T) \, P_l(\cos \tilde \theta) \,, 
\\
\Phi^p_L(E_1, E_2)&=&\frac{G_F^2{\sum_f (C_V^f)}^2}{2\alpha}
        \, Z_L(k) \, n_B(\omega_L,T) \, 
        \frac{(\omega_L^2-{k}^2)^2}{\omega_L}\frac{ E_1 E_2 
        (1-\cos^2\tilde \theta)}{E_1^2+E_2^2+2E_1E_2 \cos\tilde\theta} 
\nonumber\\ 
        && \times \quad J_L(E_1, E_2) \, \Theta(\mathcal{K}) \, 
\Theta(4E_1E_2-(\omega_L^2-k^2)) \, P_l(\cos \tilde \theta)\,, \label{phiL}
\\
\Phi^p_A(E_1, E_2)&=&\frac{G_F^2\sum_f{(C_A^f)}^2}{2\alpha} \, Z_T(k) \,
    n_B(\omega_T,T) \, \frac{\Pi_A^2(\omega_T, k)}{\omega_T E_1 E_2}
    \bigg[E_1E_2 - \frac{(k^2+E_1^2-E_2^2)(k^2-E_1^2+E_2^2)}{4k^2}
        \bigg] 
\nonumber \\
    && \times \quad J_T(E_1, E_2) \, \Theta(4E_1E_2-\Pi_T) \,  
P_l(\cos \tilde \theta)\,,\label{phiA} \\ 
\Phi^p_M(E_1, E_2)&=&\frac{G_F^2 
    \sum_f C_A^f C_V^f}{2\alpha} \, Z_T(k) \,
    n_B(\omega_T,T) \, \frac{\Pi_A(\omega_T, k) \Pi_T^2(\omega_T, k)}{k \omega_T E_1 E_2} (E_2-E_1)
     \nonumber \\ 
    && \times \quad J_T(E_1, E_2)\, \Theta(4E_1E_2-\Pi_T)\,
	 P_l(\cos \tilde \theta)\label{phiM}
\end{eqnarray}
The corresponding absorption counterparts are 
\begin{eqnarray}
R^{a}_x(E_1, E_2, \cos\theta)&=&
\frac{1-n_B(\omega, T)}{\xi n_B(\omega, T)}R^{p}_x(E_1, E_2, \cos\theta)\\
\Phi^{a}_x(E_1, E_2)&=&\frac{1-n_B(\omega, T)}{\xi n_B(\omega, T)}\Phi^{p}_x(E_1, E_2) \,,
\end{eqnarray}
where $x$ stands for $T$, $L$, $A$, or $M$.  As for the differential
emissivities in Sec. \ref{sec:DIFFEMI}, the appropriate dispersion
relations have to be solved iteratively for every choice of variables.

\subsection*{Typical Neutrino Energies}

The ratio of the total emissivity to the total rate, $Q/\Gamma$, is
a good measure of the mean neutrino plus anti-neutrino energy. Since
the dominant contribution arises from the transverse photons, we focus on    
\begin{eqnarray}\label{avgen}
\langle E_{\nu\bar{\nu}} \rangle_T = 
\frac{Q_T}{\Gamma_T} = \frac{\displaystyle
\int_0^\infty dk \, k^2\, Z_T(k) \, n_B(\omega_T,T)\,  (\omega_T^2-k^2)^3}
{\displaystyle\int_0^\infty dk \, 
{\displaystyle \frac{k^2}{\omega_T}} \, Z_T(k) \,  
n_B(\omega_T,T) \, (\omega_T^2-k^2)^3}
\end{eqnarray}
results for which are shown by the solid curves in Fig. \ref{EAVG}.
The behavior of $\langle E_{\nu \bar{\nu}} \rangle_T$ with density and
temperature can be understood by examining limiting cases.  Setting
$Z_T=1$ and using the approximate relation $\omega_T^2 - k^2
\simeq m_t^2 \approx \omega_p^2$ for semi-quantitative estimates, we
obtain
\begin{eqnarray}\label{avgen2}
\langle E_{\nu\bar{\nu}} \rangle_T \simeq T~ 
\frac { {\displaystyle \sum_{j=1}^\infty }K_2(jy)/j } 
{{\displaystyle \sum_{j=1}^\infty} K_1(jy)/(jy)} \,,  
\end{eqnarray}
where $y=\omega_p/T$ and $K_\nu$ is the modified Bessel function of
order $\nu$.  Eq. (\ref{avgen2}) can be used to obtain numerical
estimates in the extreme relativistic and nonrelativistic cases: 
\begin{eqnarray}
\langle E_{\nu\bar{\nu}} \rangle_T &\simeq& 1.46~T \quad {\rm for} \quad T \gg
\omega_p \nonumber \\ 
&\simeq& \omega_p  \,\,\,\,\, \qquad {\rm for} \quad T \ll \omega_p \,.
\end{eqnarray}
These results give a semi-quantitative account of the exact results
shown in Fig. \ref{EAVG}.  The constancy of $\langle E_{\nu \bar{\nu}} \rangle_T$
at low density and the power-law rise at high density are chiefly due
to a similar behavior of $\omega_p$ with density (see
Fig. \ref{WPMUFIGURE}).

\begin{figure}[ht]
\includegraphics[width=0.45\textwidth]{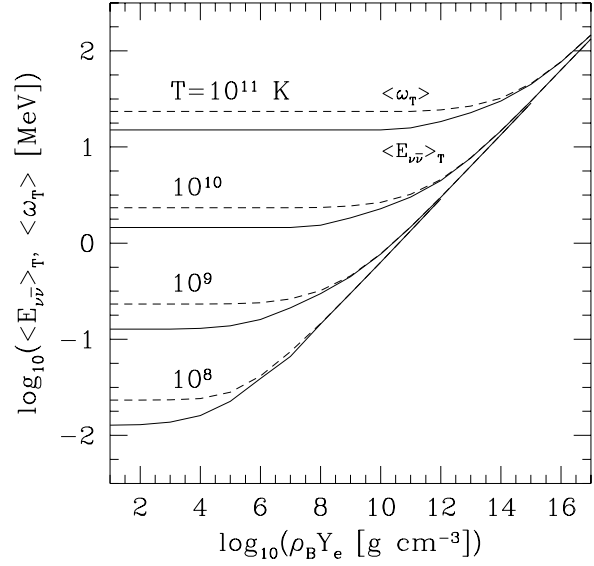}
\caption{Average transverse photon and neutrino energies
$\langle\omega_T\rangle$ and $\langle E_{\nu\bar{\nu}} \rangle_T$ as
functions of density and temperature.}
\label{EAVG}
\end{figure}

Since neutrinos stem from the decay of transverse photons in the
plasma, additional insights into the magnitude of $\langle E_{\nu \bar{\nu}}
\rangle_T$ can be obtained by examining the mean transverse photon
energies $\langle \omega_T \rangle$ as functions of density and
temperature.  Explicitly, 
\begin{eqnarray}
\langle\omega_T\rangle &=& 
\frac{\displaystyle\int_0^\infty dk~k^2 \omega_T \, n_B(\omega_T,T)}
{\displaystyle\int_0^\infty dk \, k^2 n_B(\omega_T,T)} \; \simeq \; 
 \frac{\omega_p^2}{8 T} \;
\frac { {\displaystyle\sum_{j=1}^\infty }{\displaystyle\biggl[}K_4(jy) -
K_0(jy){\displaystyle\biggr]} } 
{ {\displaystyle\sum_{j=1}^\infty }K_2(jy)/j}~ \,,  
\end{eqnarray}
where the rightmost relation is obtained by using the approximate
relation $\omega_T^2 - k^2 \simeq m_t^2 \approx \omega_p^2$.  In the
extreme relativistic and nonrelativistic cases, $\langle \omega_T
\rangle$ has the familiar forms
\begin{eqnarray}
\langle \omega_T \rangle &\simeq& 2.7~T \qquad \quad {\rm for} \quad T \gg
\omega_p \nonumber \\ 
&\simeq& \omega_p + \frac 32 ~T  \quad \, {\rm for} \quad T \ll \omega_p \,.
\end{eqnarray}
Results  obtained by using the exact dispersion relation (the dashed
curves in Fig. \ref{EAVG}) are matched well by these estimates.  As
expected, $\langle E_{\nu\bar{\nu}} \rangle_T \rightarrow \langle \omega_T
\rangle$ for $T \ll \omega_p$.  The suppression of $\langle E_{\nu\bar{\nu}}
\rangle_T$ relative to $\langle \omega_T \rangle$ for $T\gg\omega_p$ at low
density is principally due
to the suppression inherent in the squared matrix element for 
neutrino pair production from the transverse photon in the plasma.

\section{Comparison with Competing Processes}
\label{sec:Comparison}

In addition to the plasma process considered in this work,
$\nu\bar\nu$ pairs are produced from the annihilation of $e^+e^-$
pairs $(e^++e^-\rightarrow \nu+\bar{\nu})$, the interaction of electrons with 
photons  $(e^-+\gamma\rightarrow e^-+\nu+\bar{\nu})$, and the
bremsstrahlung process involving nucleons 
$(n+n\rightarrow n+n+\nu+\bar{\nu})$ . In what follows, we wish to
assess the relative importance of these processes in the ranges of
density and temperature considered in this work.  In view of the
developments reported in Ref. \cite{BRAATEN1} and this work, our
results represent an update of similar comparisons made earlier in the
literature \cite{BEAUDET1,BEAUDET2,DICUS1,SCHINDER1,BRUENN1,Haft}.
Numerical results for emissivities from the pair process are taken
from Ref. \cite{BRUENN1}, the photo process from Ref. \cite{SCHINDER1} 
and the bremsstrahlung process from Ref. \cite{HR98}. 
\begin{figure}[ht]
\includegraphics[width=0.45\textwidth]{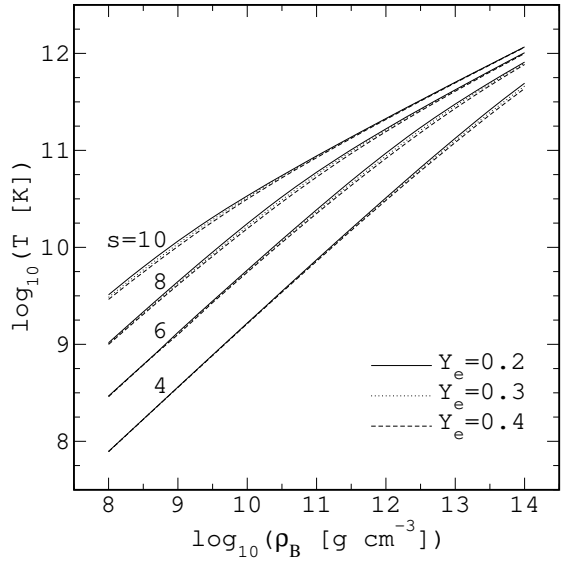}
\caption{Profiles of temperature versus baryon density $\rho_B$ for
varying values of the electron fraction $Y_e = n_e/n_B$, at
constant entropies per baryon $s$.}
\label{TPROFS}
\end{figure}

\begin{figure*}
\includegraphics[width=0.95\textwidth]{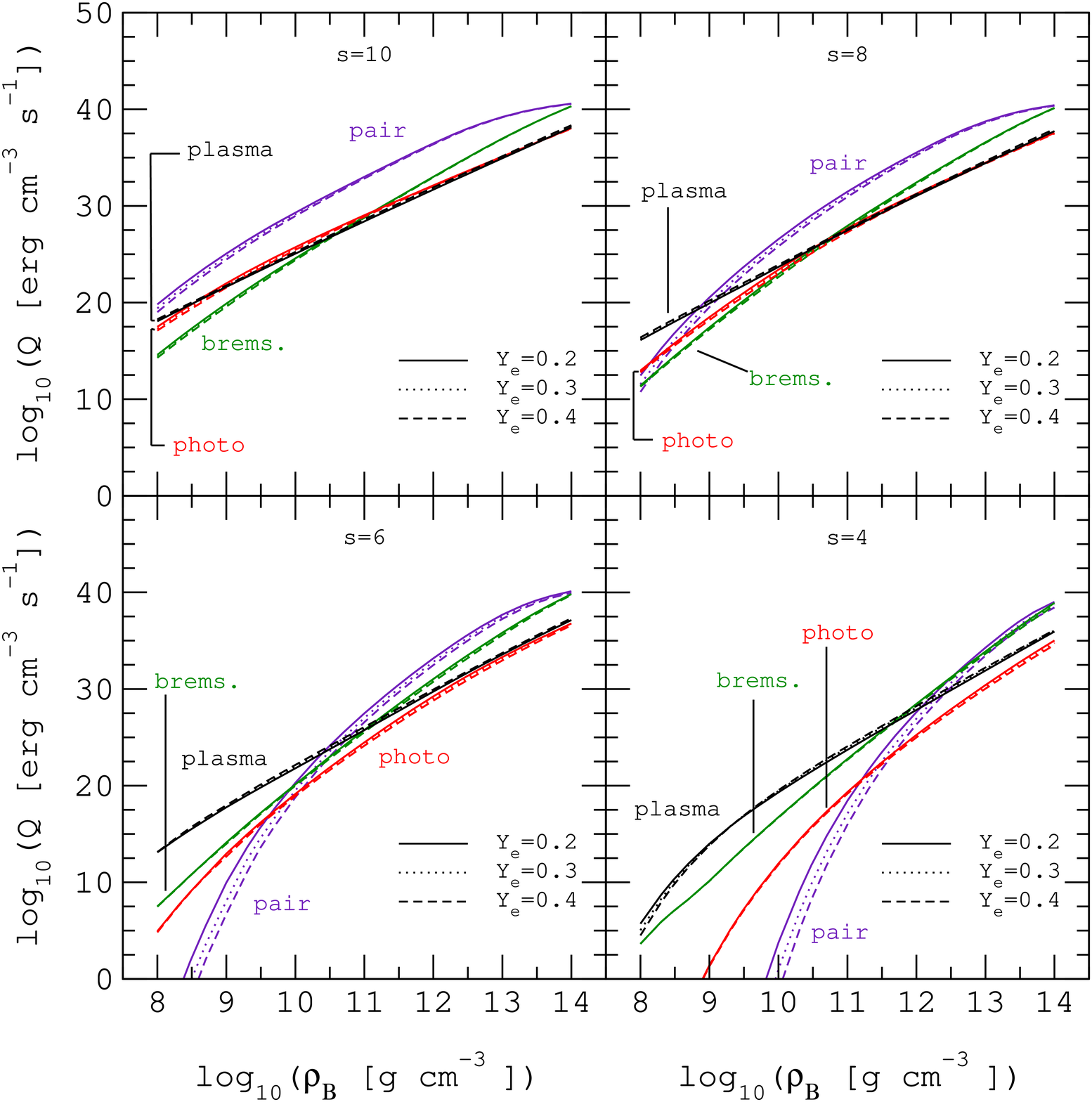}
\caption{Comparison of neutrino emissivities from the pair, plasma,
photo, and bremsstrahlung processes at constant entropies per baryon
of $s = 4, 6, 8$, and 10, respectively.  The corresponding
temperatures are shown in Fig.~\ref{TPROFS}.  }
\label{S46810}
\end{figure*}

Traditionally, comparisons of total emissivities from the various
processes have been carried out using $T$ and $\rho_BY_e$ as
variables.  In the context of supernova dynamics, additional insights
can be gained by making comparisons along constant entropy per baryon
adiabats as a function of baryon density $\rho_B$.  We therefore
present constant entropy per baryon adiabats in Fig. \ref{TPROFS}. The
results in this figure are based on calculations of the equation of
state of matter in which the net negative electric charge of electrons
and positrons is cancelled by a uniform positively charged background
of protons, alpha particles, and heavier ions (see, for example,
Ref. \cite{LS} and references therein).  The phase structure of matter
and the relative abundances of the various constituents including
those of dripped neutrons at sub-nuclear densities are determined by
the minimization of free energy.

In Fig. \ref{S46810}, we compare the total neutrino emissivities from
the various processes at constant entropy per baryon (in units of
$k_B$) of $s=4,6,8$ and 10, respectively.  The plasma neutrino
emissivity dominates over the other processes considered for entropies
and densities that lie at the lower ends of those shown in this
figure. Toward the higher ends of entropies and densities, the pair
and bremsstrahlung processes are far more efficient than both the
photo and plasma processes. Once the threshold of $T \sim 2m_e$ is
crossed, $e^+e^-$ annihilation becomes the dominant source of
$\nu\bar\nu$ pairs, the emissivity growing as $T^9$.  The dominance of
the bremsstrahlung process at low entropies and high densities is due
to the increasing number of participating Pauli-unblocked nucleon
pairs that can benefit from strong interaction (pion-exchange)
dynamics. Note that the electron
fraction $Y_e$ has a significant effect only for the pair emissivity,
which depends directly on the number of $e^+e^-$ pairs, which grows
with increasing $s$ (or $T$ at a given density).  The plasma and
photo processes are much less sensitive to $Y_e$, primarily because of
the presence of electrons even without positrons.  The insensitivity
of the bremsstrahlung process to $Y_e$ is due to the preponderance 
of $nn$ pairs under these physical conditions. 

The relative importance of the plasma process depends on whether or
not favorable conditions are encountered during the course of
supernova dynamics.  Simulations peformed to date
\cite{Bruenn93,Janka03} point to the low density ($\rho_B <
10^{11}~{\rm g~cm^{-3}}$) and low entropy ($s < 6$) regions encountered
a few tens of milliseconds after bounce.  For electron neutrinos and
antineutrinos, the pair reactions are only corrections compared to the
beta decay and capture processes. For $\mu$  and $\tau$  neutrinos,
densities of $\rho_B < 10^{11}~{\rm g~cm^{-3}}$ lie below the
neutrinosphere located at $\rho_B \sim 10^{12}~{\rm g~cm^{-3}}$. The largest
pair rate at the latter density is 5-10 orders larger than the plasma
rate in its most relevant region. At $s =4$, plasma neutrinos are,
however, competitive with those from pair and bremsstrahlung processes
at the neutrinospheric density (but such low entropies in current post
bounce models occur only at much higher densities).  In order to judge
the relative importance for neutrino transport, consult the discussion
of thermalization depths for the different processes in
Ref.~\cite{KRJ02}.  It must be emphasized, however, that the values of
$s, T$ and $\rho_B$ attained in such simulations are predicated both
by the input microscopic physics and the complex interplay of neutrino
transport with hydrodynamics (in many cases coupled with convection,
rotation and magnetic fields), both of which are being
currently improved.  The differential emissivities presented in this
work are a part of such improvements in the area of microphysical
ingredients and their relative importance remains to be ascertained in  
the future. 

\section{Summary and Outlook}
\label{sec:summary}

In summary, 

\begin{itemize}

\item We have calculated the differential rates and emissivities of
neutrino pairs from an equilibrium plasma for baryon densities $\rho_B
< 10^{14}~{\rm g~cm^{-3}}$ and temperatures $T \le 10^{11}$ K.  We
have checked that the new analytical expressions for the differential
emissivities yield total emissivities that are consistent with those
calculated using independent methods.

\item We have developed new analytical expressions for the total
emissivities in various limiting situations.  These results help us to
better understand qualitatively the scaling of the results with
physical quantities such as the chemical potential and plasma
frequency at each temperature and density.  A comparison with other
competing processes, such as $e^+e^-$ annihilation, $e^-\gamma$
interaction, and nucleon-nucleon bremsstrahlung, shows that the plasma
process is the dominant source of neutrino pairs in regions of high
degeneracy.

\item Using our results for the differential rates and emissivities,
we have calculated the production and absorption kernels in the source
term of the Boltzmann equation employed in exact, albeit numerical,
treatments of multi-energy neutrino transport.  We have also provided
the appropriate Legendre coefficients of these kernels in forms
suitable for multi-group flux-limited diffusion schemes.

\end{itemize}

The significance of neutrino pair production from the plasma process
in detailed calculations of core-collapse supernovae in which neutrino
transport is strongly coupled with hydrodynamics remains to be
explored.  The detailed differential information provided in this work
may also be of utility in better understanding the evolution of other
astrophysical systems such as the red giant stages of stellar
evolution, the cooling and accretion-induced collapse of white dwarfs,
burning phenomena atop neutron stars, the explosive stages of type Ia
supernovae, and gamma-ray bursters.

\begin{acknowledgments}
We thank Doug Swesty, Jim Lattimer, and Eric Myra who alerted us to
the need for differential rates and emissivities in simulations of
neutrino transport in the supernova environment.  Special thanks are
due to Jim Lattimer for useful suggestions and for a careful reading
of the manuscript.  We are grateful to Steve Bruenn and Thomas Janka
for helpful discussions and for providing us with relevant details
from their supernova simulations.  The work of S.R. and M.P. was
supported by the US-DOE grant DE-FG02-88ER40388 and that of S.I.D. by
the National Science Foundation Grant No. 0070998. Travel support for
all three authors under the cooperative agreement DE-FC02-01ER41185
for the SciDaC project ``Shedding New Light on Exploding Stars:
Terascale Simulations of Neutrino-Driven Supernovae and Their
Nucleosynthesis'' is gratefully acknowledged.

\end{acknowledgments}

\appendix
\section{Effective  Coupling}
\label{sec:EFC}
The decay of a massive photon or plasmon into a $\nu\bar{\nu}$ pair
requires the coupling of the intermediate off-shell $e^+e^-$ pair to
the outgoing neutrinos. Since neutrinos can belong to any of the three
flavors, and the loop calculation is performed including the $e^+e^-$
pair only, different $\nu$ flavors will have different rates of
emission due to different couplings to the $e^+e^-$ pair.

The electron-positron couplings to the neutrino-antineutrino pair
described by the two one-loop Feynman diagrams in Fig. \ref{pair} are
given by the matrix elements:
\begin{eqnarray}
\mathcal{M}_Z&=&\left(\bar{u_{\nu}}\frac{-ig_Z}{2}\gamma^{\mu}
        (c_V^i-c_A^i\gamma_5)v_{\nu}\right) (-i) 
        \frac{g_{\mu\nu}-Q_{\mu}Q_{\nu}/M_Z^2}{Q^2-M_Z^2}
        \left(\bar{v_{e}}\frac{-ig_Z}{2}\gamma^{\nu}
        (c_V^e-c_A^e\gamma_5)u_{e}\right)\,,
\label{MZ}\\
\mathcal{M}_W&=&\left(\bar{u_{\nu}}\frac{-ig_W}{2\sqrt{2}}\gamma^{\mu}
        (1-\gamma_5)u_{e}\right)
        (-i)\frac{g_{\mu\nu}-Q_{\mu}Q_{\nu}/M_W^2}{Q^2-M_W^2}
        \left(\bar{v_{e}}
\frac{-ig_W}{2\sqrt{2}}\gamma^{\nu}(1-\gamma_5)v_{\nu}\right)\,, 
\label{MW}
\end{eqnarray}
where in Eq.~(\ref{MZ}) the superscript $i$ denotes  neutrino
flavor.  The coefficients $c_V$ and $c_A$ are: 
\begin{eqnarray}
c_V&=&\frac 12 \qquad {\rm and}~~ c_A=\frac 12 \qquad{\rm for}~~ \nu_e,~
\nu_{\mu}~~{\rm and}~~\nu_{\tau} \nonumber \\
c_V&=&-\frac{1}{2}+2\sin^2\theta_W~~{\rm and}~~
c_A=-\frac{1}{2}~~{\rm for}~~ e,~\mu, 
~{\rm and}~\tau\,.
\end{eqnarray}
While the exchange of the $Z-$boson produces all three neutrino and 
antineutrino flavors ($\nu_{e,\mu,\tau}$), $W-$boson exchange produces
only $\nu_e \bar{\nu}_e$ pair. This gives rise to different production 
rates for different $\nu$ types so that the electron neutrino will have
a different emissivity compared to those of the $\nu_{\mu}$ and $\nu_\tau$.

It is economical to rephrase these two diagrams into an effective form
dependent on the $\nu$ type. After performing Fierz rearrangement and
discarding the $Z$ and $W$ momenta ($M_{Z,W}^2 \gg q^2$), we obtain
the effective $e^+e^-$ coupling
\begin{eqnarray}
\mathcal{M}_{eff}&=&\frac{-iG_F}{\sqrt{2}}
\big(\bar{u_{\nu}}\gamma^{\mu}(1-\gamma_5)
        v_{\nu}\big) \label{Meff}
        \big(\bar{v_{e}}\gamma_{\mu}(C_V^f-C_A^f\gamma_5)u_{e}\big) \,, 
\label{effectiveenu}
\end{eqnarray}
where the couplings $C_V^f$ and $C_A^f$ now depend on the outgoing
neutrino flavor. Their numerical values are
\begin{eqnarray}
C_V&=&\frac{1}{2}+2\sin^2\theta_W,\quad C_A=\frac{1}{2}\quad {\rm for}~
\nu_e \nonumber \\
C_V&=&-\frac{1}{2}+2\sin^2\theta_W,\quad C_A=-\frac{1}{2}\quad {\rm
for}~ \nu_\mu~{\rm and}~ \nu_\tau\,.
\end{eqnarray}
The reduction of the interaction to a single
effective interaction significantly simplifies calculations of rates
and emissivities from the process $\gamma^*\rightarrow\nu+\bar{\nu}$.





\begin{thebibliography}{99}
\bibitem{fnote}
The polarization vector $\epsilon^{(3)\mu}$ in
Eqs. (\ref{epsilon}) and (\ref{Afinal})
is a 4--vector introduced as a part of the orthogonal
set $\{K^{\mu},\, \epsilon^{(1)\mu},\, \epsilon^{(2)\mu},\,
\epsilon^{(3)\mu}\}$. The frequency $\omega$ and wave vector $k$ 
occuring in $\epsilon^{(3)\mu}$  satisfy the transverse dispersion relation.
\end{thebibliography}
\end{document}